\documentstyle[12pt,cite]{article}
\catcode`\@=11
\renewcommand{\thefootnote}{\fnsymbol{footnote}}

\begin{document}

\begin{titlepage}

Int. J. Mod. Phys. {\bf A10} (1995) 289 -- 336.





{\centerline{\large \bf UNIFORMIZATION THEORY AND 2D GRAVITY}}


{\centerline{\large \bf I. LIOUVILLE ACTION AND INTERSECTION NUMBERS}}

\vspace{0.1cm}

\centerline{\large{\sc Marco}  {\sc  Matone}\footnote{e-mail: 
matone@padova.infn.it, mvxpd5::matone}}


\centerline{\it Department of Mathematics}
\centerline{\it Imperial College}
\centerline{\it 180 Queen's Gate, London SW7 2BZ, U.K.}

\centerline{\it and}

\centerline{\it Department of Physics ``G. Galilei'' - Istituto Nazionale di 
Fisica Nucleare}
\centerline{\it University of Padova}
\centerline{\it Via Marzolo, 8 - 35131 Padova,
Italy\footnote[5]{Present address}}

\vspace{0.1cm}

\centerline{\large ABSTRACT}


This is the first part of an investigation concerning the formulation of 2D 
gravity in the framework of the uniformization theory of Riemann surfaces. 
As a first step in this direction we show that the classical Liouville action 
appears in the expression of the correlators of topological gravity.
Next we derive an inequality involving the cutoff of 2D gravity and the 
background geometry. Another result, still related to uniformization theory, 
concerns a relation between the higher genus normal ordering and the Liouville 
action. We introduce operators covariantized by means of the inverse map 
of uniformization.
These operators have interesting properties including holomorphicity.
In particular they are crucial to show that
the chirally split anomaly of CFT is equivalent to the Krichever-Novikov cocycle
and vanishes for deformation of the complex structure induced by the
harmonic Beltrami differentials.
By means of the inverse map we propose a realization of the 
Virasoro algebra on arbitrary Riemann surfaces and find the eigenfunctions for 
the {\it holomorphic} covariant operators defining higher order cocycles
and anomalies which are related to $W$-algebras. Finally 
we face the problem of considering the positivity of $e^\sigma$, with 
$\sigma$ the Liouville field, by proposing an explicit construction for the 
Fourier modes on compact Riemann surfaces.
These functions, whose underlying number theoretic structure
seems related to Fuchsian groups and to the eigenvalues of the Laplacian,
are quite basic and may provide the building blocks to
properly investigate the long-standing uniformization problem
posed by Klein, Koebe and  Poincar\'e.

\end{titlepage}

\newpage
\tableofcontents

\setcounter{footnote}{0}

\renewcommand{\thefootnote}{\arabic{footnote}}

\renewcommand{\theequation}{\thesection.\arabic{equation}}
\newcommand{\mysection}[1]{\setcounter{equation}{0}\section{#1}}

\mysection{Introduction}

The last few years have witnessed remarkable progress in 2D quantum gravity 
\cite{smnl,lg,ddk,JLG,mm,1}. Important connections between non-perturbative 2D 
gravity, 
generalized KdV hierarchies, topological and Liouville gravity have been 
discovered \cite{2,wd,kont}  (see \cite{mm2} for
reviews).
This interplay has provided a good opportunity for important 
progress both in  physics and mathematics.

Despite these results there are still important unsolved problems. For 
example a direct and satisfactory proof of the equivalence of the different 
models of 2D quantum gravity is still lacking. In particular in the continuum 
formulation of 2D gravity the higher genus correlators have not yet been 
worked out. One of the problems concerns the integration on the moduli space.
The Schottky problem hinders integration on the 
Siegel upper plane. Analogously to the relations between intersection 
theory on the moduli space and KdV \cite{2,wd,kont}, 
the solution of the Schottky problem is based on deep 
relationships between algebraic geometry and integrable systems. In 
particular Shiota and Mulase \cite{sm} proved that, according to the Novikov 
conjecture, a matrix in the Siegel upper plane is a Riemann period matrix if 
and only if the corresponding $\tau$-function satisfies Hirota's bilinear 
relations. Apparently it is technically
impossible to satisfy these constraints in performing
the integration on the Siegel upper plane.  Similar aspects are intimately 
linked with 2D gravity. In particular, results from matrix models, where 
the integration on the moduli space is implicitly 
performed, suggest that at least in some cases the integrand is a total 
derivative.

Many of the results concerning the continuum formulation of 
2D gravity have been derived by considering
its formulation in Minkowskian space whereas Liouville theory is
intrinsically Euclidean. Furthermore, most of the results are known on the 
sphere and the torus. On the other hand the Liouville equation is the 
condition of constant negative curvature. Thus to get some insight into
Liouville theory it is necessary to concentrate the analysis of the 
continuum formulation of 2D gravity on surfaces with negative Euler 
characteristic. A crucial step in setting the mathematical formalism for this 
purpose has been made by Zograf and Takhtajan \cite{0,asym}. Starting with 
the intention of proving a conjecture by Polyakov, they showed that Liouville 
theory is strictly related to uniformization theory of Riemann surfaces. 
In particular it turns out that the Liouville action evaluated on the classical 
solution is the K${\rm \ddot a}$hler potential for the Weil-Petersson 
metric on the Schottky space. It seems that the results in \cite{0,asym} 
have relevant implications for 2D gravity which are still 
largely unexplored. In our 
opinion they will serve as a catalyst to formulate a ``quantum 
geometrical'' 
approach to 2D gravity in the framework of uniformization theory. 
Some aspects concerning uniformization theory and 2D gravity have been 
considered in \cite{tt1}.

The present paper is the first part of a work whose basic aim is to attempt 
to bring together the different branches of mathematical technology 
in order to investigate 2D 
gravity in a purely geometrical context.

The organization of the paper is as follows. In section 2 we summarize 
basic facts about the uniformization theory of Riemann surfaces
and the Liouville equation. The points we will consider include the 
explicit construction of differentials in terms of theta functions. This 
is particularly useful because by means of essentially two theorems it is 
possible to recover and understand,
by some pedagogical ``theta gymnastics'', a lot of basic facts concerning
 Riemann surfaces. Furthermore, we will discuss in detail the properties
of the Poincar\'e metric in relation to Liouville equation and 
uniformization theory. One of the aims of this section is to clarify
some points concerning the properties
of the Liouville field. For example note that
in current literature the field $e^{\gamma\sigma}$ is sometimes considered
as a $(1,1)$-differential, which is in contradiction
with the fact that, since $g=e^{\gamma\sigma}\hat g$, 
with $\hat g$ a background 
metric, $e^{\gamma\sigma}$ must be a $(0,0)$-differential. 
This point is related to the difficulties arising in the definition
of conformal weights in Liouville theory. 

In section 3 we introduce a set of operators which are 
covariantized by means of the inverse of the uniformization map and 
give their chiral (polymorphic) and non chiral eigenfunctions.
We will see that these operators have important properties including
holomorphicity. 
Next we consider the cocycles associated to the above operators.
In particular, the cocycle associated to the covariantized third
derivative is the Fuchsian form of the Krichever-Novikov 
($KN$) cocycle. In this framework we show that the normal ordering for 
operators defined on $\Sigma$ is related to {\it classical} Liouville theory.
Another result concerns the equivalence between the chirally split anomaly
of CFT and the $KN$ cocycle. 
In particular it turns out that this anomaly vanishes under deformation 
of the complex structure induced by the harmonic Beltrami differentials.
These results suggest to consider higher order anomalies 
in the framework of uniformization theory of 
vector bundles on Riemann surfaces (${W}$-algebras).

In section 4 we consider a sort of higher genus generalization of the Killing
vectors which is based on the properties of the Poincar\'e metric and of the 
inverse map of uniformization. This analysis will suggest a realization of 
Virasoro algebra on arbitrary Riemann surfaces.

In section 5 we introduce an infinite  set of regular functions which can be 
considered as ``building-blocks'' to develop the higher genus Fourier analysis. 
We argue that the number theoretic structure
underlying the building-blocks
is related to the uniformization problem. In particular we investigate 
the structure of the eigenvalues of the Laplacian. 
Besides its mathematical interest one of the aims
 of this investigation is to provide a suitable tool to recognize the 
modes of the Liouville field $\sigma$. This is an 
attempt to face the  problem, usually untouched in the literature,
of considering metric positivity in performing the quantization
of Liouville theory.

In section 6 we find a direct link between Liouville and
topological gravity.
In particular we will show that
the first tautological class, which enters in the 
correlators of topological gravity \cite{1,2,wd}, has the classical Liouville 
action as potential, in particular
\begin{equation} 
\kappa_1={i\over 2\pi^2}\overline\partial\partial S_{cl}^{(h)}.
\label{000}\end{equation}

In the last section we discuss important aspects concerning
the role of the Poincar\'e metric
chosen as background. Furthermore, by means of classical results on univalent 
functions, we
derive an interesting inequality involving the cutoff of 2D gravity and the 
background geometry whose consequences should be further investigated.

\mysection{Uniformization Theory And Liouville Equation}

In this section we introduce background material for later use.
In particular we begin by giving a procedure to explicitly
construct any meromorphic differential defined on a Riemann surface.
After that we introduce basic facts concerning the uniformization theory 
\cite{fk1,kra0} and the Liouville equation.
Then we investigate the properties of the inverse  map 
of uniformization, and consider the linearized version of the Schwarzian 
equation $\{J_H^{-1},z\}=T^F$, with $T^F$ the Liouville stress tensor
(or Fuchsian projective connection).
We conclude the section with some remarks on the standard approach
to Liouville gravity.

\subsection{Differentials On $\Sigma$: Explicit Construction}

Let us start by recalling some basic facts about the space of the 
$(p,q)$-differentials ${\cal T}^{p,q}$. 
Let $\{(U_\alpha,z_\alpha)|\alpha\in I\}$ be an atlas 
with harmonic coordinates on a Riemann surface $\Sigma$.
A differential in ${\cal T}^{p,q}$ is a set 
of functions $f\equiv\{f_\alpha(z_\alpha,\bar z_\alpha)|\alpha\in I\}$
where each $f_\alpha$ is defined on $U_\alpha$. These functions are 
related by the following transformation
in $U_\alpha \cap U_\beta$
\begin{equation}
f_\alpha(z_\alpha,\bar z_\alpha)(dz_\alpha)^p(d\bar z_\alpha)^q=
f_\beta(z_\beta,\bar z_\beta)(dz_\beta)^p(d\bar z_\beta)^q,
\qquad f\in {\cal T}^{p,q},
\label{hU}\end{equation}
that is $f_\alpha$ transforms as 
$\partial_{z_\alpha}^p\partial_{\bar z_\alpha}^q$.

In the case of the Riemann sphere 
$\widehat {\bf C}\equiv {\bf C}\cup\{\infty\}$,
all the possible transition functions $z_-=g_{-+}(z_+)$ between
the two patches $(U_\pm,z_\pm)$ of the standard  atlas
are holomorphically equivalent to $g_{-+}(z_+)=z_+^{-1}$, that is
$\widehat {\bf C}$ has one complex structure only (no moduli).
Therefore giving $f_+(z_+,\bar z_+)$ fixes
$f_-(z_-,\bar z_-)$ and vice versa.

In the higher genus case fixing a component of $f$ in a patch 
is not sufficient to uniquely fix the other functions  in 
$f\equiv\{f_\alpha(z_\alpha,\bar z_\alpha)|\alpha\in I\}$.
As an example we consider the case
of meromorphic $n$-differentials $f^{(n)}$
on a compact Riemann surface of genus $h$.
The Riemann-Roch theorem guarantees that 
it is possible to fix
the points in\footnote{Let
$\{P_k\}$ ($\{Q_k\}$) be
 the set of zeroes (poles) of a $n$-differential $f^{(n)}$.
The formal sum ${\rm Div}\, f^{(n)}\equiv \sum_{k=1}^pP_k-\sum_{k=1}^qQ_k$
and ${\rm deg}\,f^{(n)}\equiv p-q$,
define the divisor and the degree of $f^{(n)}$ respectively.
It turns out that ${\rm deg}\,f^{(n)}=2n(h-1)$.}  ${\rm Div}\, f^{(n)}$
 up to (in general) $h$ zeroes, say $P_1,\ldots,P_h$,
whose position is fixed by $P_{h+1},\ldots,P_p,Q_1,\ldots,Q_q$,
$q=p-2n(h-1)$, the conformal structure of $\Sigma$
and (in general) on the choice of the local coordinates.
An instructive way to see this is to explicitly construct
$f^{(n)}$.
In order to do this
we first recall some facts about theta functions.
Let us denote by $\Omega$ the $\beta$-period matrix 
\begin{equation}
\Omega_{jk}\equiv 
\oint_{\beta_j}\omega_k,\label{jwelmk}\end{equation}
where $\omega_1,\ldots,\omega_h$ are the 
holomorphic differentials
with the standard
normalization
\begin{equation}
\oint_{\alpha_j}\omega_k=\delta_{jk},
\label{stndnorm}\end{equation}
$\alpha_k,\beta_k$ being the homology cycles basis.
The theta function with characteristic reads
\begin{equation}
\Theta \left[^a_b\right]\left(z|\Omega\right)=
\sum_{k\in {\bf Z}^h}e^{\pi i (k+a)\cdot\Omega\cdot(k+a)+
2\pi i  (k+a)\cdot (z+b)}, 
\qquad\Theta\left(z|\Omega\right)\equiv\Theta 
\left[^0_0\right]\left(z|\Omega\right),
\label{oixho}\end{equation}
where $z\in{\bf C}^h,\; a,b\in{\bf R}^h$.
When $a_k,b_k\in\{0,1/2\}$,  $\Theta \left[^a_b\right]
\left(-z|\Omega\right)=(-1)^{4a\cdot b}\Theta \left[^a_b\right]
\left(z|\Omega\right)$. The $\Theta$-function
is multivalued under a lattice shift in the $z$-variable
\begin{equation}
\Theta \left[^a_b\right]\left(z+n+\Omega\cdot m|\Omega\right)=
e^{-\pi i m \cdot\Omega\cdot m -2\pi i m \cdot z
+2\pi i (a\cdot n-b\cdot m)}
\Theta\left[^a_b\right]\left(z|\Omega\right).
\label{ojcslc}\end{equation}
Let us now introduce the prime form
$E(z,w)$. It is a holomorphic
$-1/2$-differential both in $z$ and $w$,
vanishing for $z=w$ only
\begin{equation}
E(z,w)={\Theta\left[^a_b\right]\left(I(z)-I(w)|\Omega\right)\over
h(z)h(w)}. \label{pojdlk}\end{equation}
Here $h(z)$ denotes
the square root of $\sum_{k=1}^h\omega_k(z)
\partial_{u_k}\Theta\left[^a_b\right]\left(u|\Omega\right)|_{u_k=0}$;
it is the holomorphic 1/2-differential with non singular 
(i.e. $\partial_{u_k}\Theta\left[^a_b\right]
\left(u|\Omega\right)|_{u_k=0}\ne 0$) odd spin structure
$\left[^a_b\right]$.
The function $I(z)$ in (\ref{pojdlk}) 
denotes the Jacobi map 
\begin{equation}
I_k(z)=\int_{P_0}^z\omega_k,\qquad z\in\Sigma,
\label{ghftdt}\end{equation}
with $P_0\in \Sigma$ an arbitrary base point. 
This map is an embedding of $\Sigma$ into
the Jacobian 
\begin{equation}
J(\Sigma)={\bf C}^h/L_\Omega,\qquad L_\Omega={\bf Z}^h
+\Omega {\bf Z}^h.\label{jac}\end{equation}
By (\ref{ojcslc}) it follows that the multivaluedness of $E(z,w)$ is
\begin{equation}
E(z+{n}\cdot {\alpha} +{m}\cdot {\beta},z)=
e^{-\pi i m\cdot \Omega\cdot m -2\pi i m\cdot
\left(I(z)-I(w)\right)}E(z,w).
\label{primeform}\end{equation}
In terms of $E(z,w)$ one can construct the following
$h/2$-differential with empty divisor
\begin{equation}
\sigma(z)=\exp\left(-\sum_{k=1}^h\oint_{\alpha_k}\omega_k(w)\log
E(z,w)\right),\label{sigma}\end{equation}
whose multivaluedness is
\begin{equation}
\sigma(z+{n}\cdot {\alpha} +{m}\cdot {\beta})=
e^{\pi i (h-1) {m} \cdot \Omega\cdot {m}-2\pi i 
{m}\cdot \left(\Delta-(h-1){I}(z)\right)}
\sigma(z),\label{mltvld4}\end{equation}
where $\Delta$ is (essentially) the {\it vector of Riemann constants}
\cite{fay}.
Finally we quote two theorems:
\begin{itemize}
\item[{\bf a.}]{{\bf Abel Theorem} \cite{fk1}. {\it A necessary
and sufficient condition for ${\cal D}$ to be the divisor
of a meromorphic function is that}
\begin{equation}
I\left({\cal D}\right)=0\; {\rm mod}\, \left(L_\Omega\right)\;
and \; {\rm deg}\, {\cal D}=0.\label{abella}\end{equation}}
\item[{\bf b.}]{{\bf Riemann vanishing theorem} \cite{fay}.
{\it The function}
\begin{equation}
\Theta\left(I(z)-\sum_{k=1}^hI(P_k)+\Delta\bigg|\Omega\right),
\qquad z,P_k\in \Sigma,\label{rvth}\end{equation}
{\it either vanishes identically or else it has 
$h$ zeroes at $z=P_1,\ldots,P_h$}.}
\end{itemize}

We are now ready to explicitly construct the differential
$f^{(n)}$ defined above. First of all note that
\begin{equation}
\widetilde f^{(n)}=\sigma(z)^{2n-1}
{\prod_{k=h+1}^p E(z, P_k)\over
\prod_{j=1}^{p-2n(h-1)} E(z, Q_j)},\label{explkpr}\end{equation}
is a multivalued $n$-differential
with ${\rm Div}\, \widetilde f^{(n)}=
\sum_{k=h+1}^pP_k-\sum_{k=1}^{p-2n(h-1)}Q_k$. Therefore we set
\begin{equation}
f^{(n)}(z)=g(z) 
\widetilde f^{(n)},\label{explk}\end{equation}
where, up to a multiplicative constant, $g$ is fixed 
by the requirement that $f^{(n)}$ be
singlevalued. From the multivaluedness of 
the $E(z,w)$ and $\sigma(z)$
it follows that, up to a multiplicative constant
\begin{equation}
g(z)=\Theta\left(I(z)+{\cal D}
\big|\Omega\right),
\label{thetafncte}\end{equation}
with
\begin{equation}
{\cal D}=\sum_{k=h+1}^pI(P_k)-
\sum_{k=1}^{p-2n(h-1)}I(Q_k)+(1-2n)\Delta.
\label{iudlkm}\end{equation}
By Riemann vanishing theorem $g(z)$ has just $h$-zeroes 
$P_1,\ldots,P_h$ fixed by  ${\cal D}$. Thus the
requirement of singlevaluedness also fixes the position of the 
remainder $h$ zeroes.
To make manifest the divisor in the RHS of
(\ref{explk}) we first recall that the
image of the canonical line bundle $K$ on the Jacobian
of $\Sigma$ coincides
with $2\Delta$ \cite{fay}. 
 On the other hand, since
\begin{equation}
\left[K^n\right]=
\left[ \sum_{k=1}^pP_k-\sum_{k=1}^{p-2n(h-1)}Q_k\right],
\label{abel}\end{equation}
by Abel theorem we have\footnote{The square brackets in (\ref{abel})
denote the divisor class associated to the line bundle $K^n$.
Two divisors belong to the same class if they differ by a divisor 
of a meromorphic function.}
\begin{equation}
{\rm Div}\, \Theta\left(I(z)+{\cal D}
\big|\Omega\right)={\rm Div}\, \Theta\left(I(z)-\sum_{k=1}^hI(P_k)+
\Delta\bigg|\Omega\right),
\label{thetafncte1}\end{equation}
and by Riemann vanishing theorem
\begin{equation}
{\rm Div}\, \Theta\left(I(z)+{\cal D}
\big|\Omega\right)=\sum_{k=1}^hI(P_k).
\label{thetafncte2}\end{equation}

Above we have considered harmonic coordinates. If
one starts with arbitrary coordinates
\begin{equation}
ds^2=\widetilde g_{ab}dx^adx^b,\label{gen2}\end{equation}
the harmonic ones, defining
the conformal form $2g_{z\bar z}|dz|^2$, are determined by the Beltrami
equation\footnote{In isothermal coordinates the metric reads
$ds^2=e^\phi\left((dx)^2 +(dy)^2\right)$, where $z=x+iy$.
Note that considering $x=cst$ as an isothermal curve,
$y=cst$ corresponds to the curve of heat flow.}
$\widetilde g^{1\over 2} 
\epsilon_{ac}\widetilde g^{cb}\partial_b z=i\partial_a z$. 
Therefore we can 
globally choose $ds^2=e^\phi |dz|^2$, $e^\phi=2g_{z\bar z}$.
This means that with respect to the 
new set of coordinates $\{(U_\alpha,z_\alpha)|\alpha\in I\}$ 
the metric is in the 
conformal gauge
$ds_\alpha^2=e^{\phi_\alpha} |dz_\alpha|^2$ in each patch. That is
the functions in 
$\phi\equiv
\{\phi_\alpha(z_\alpha,\bar z_\alpha)|\alpha\in I\}$ are related by 
the following transformation in $U_\alpha \cap U_\beta$
\begin{equation}
\phi_\alpha(z_\alpha,\bar z_\alpha)=\phi_\beta(z_\beta,\bar z_\beta)+
\log \big | dz_\beta/dz_\alpha\big |^2.\label{tr6}\end{equation}
By a rescaling $ds^2\to d{\widetilde s}^2=\rho ds^2$ 
it is possible to set, at least in one patch,
$d{\widetilde s}_\alpha^2=|dz_\alpha|^2$.
 Since $\rho\in {\cal T}^{0,0}$,
 there is at least one patch $(U_\gamma,z_\gamma)$
where $d{\widetilde s}_\gamma^2\ne cst|dz_\gamma|^2$.
Finally we recall that a property of the metric is positivity. Thus if 
$g=e^\sigma\hat g$ with $\hat g$ a well-defined metric, then 
$e^\sigma\in{\cal C}_+^\infty$ where
${\cal C}_+^\infty$ denotes the subspace of positive smooth
functions in ${\cal T}^{0,0}$. Later, in the framework of the 
``Liouville condition'',
we will discuss some aspects
concerning the structure of
the boundary of ${\cal C}_+^\infty$.

\subsection{Uniformization And Poincar\'e Metric}

Let us denote by $D$ either the Riemann sphere   
$\widehat{\bf C}={\bf C}\cup \{\infty\}$, the complex plane $\bf C$, or 
the upper half plane $H=\{w\in{\bf C}|{\rm Im}\,w>0\}$.
The uniformization theorem states that every Riemann surface $\Sigma$ 
is conformally equivalent to 
the quotient $D/\Gamma$ with $\Gamma$ a freely acting discontinuous
group of fractional transformations preserving $D$. 

Let us consider the case of Riemann surfaces with universal covering 
$H$ and denote by $J_H$ the complex analytic covering $J_H:H\to \Sigma$.
In this case $\Gamma$  (the automorphism group of $J_H$)
is a finitely generated  
Fuchsian group $\Gamma \subset
PSL(2,{\bf R})=SL(2,{\bf R})/\{ I,- I\}$ 
acting on $H$ by linear fractional transformations
\begin{equation}
w\in H,\qquad \gamma \cdot w=
 {aw+b\over cw+d}\in H,\;\qquad \gamma=\left(\begin{array}{c}a\\c
\end{array}\begin{array}{cc}b\\d\end{array}\right)\in\Gamma\subset
PSL(2,{\bf R}).
\label{3}\end{equation}
By the fixed point equation 
\begin{equation}    
w_\pm={a-d\pm
\sqrt { (a+d)^2-4} \over 2c}, \label{fxdpnt}\end{equation}
it follows that $\gamma\ne I$ can be classified
according to the value of $|{\rm tr}\,\gamma|$:

\begin{itemize}

\item[1.] {{\bf Elliptic.} $|{\rm tr}\, \gamma|<2$, $\gamma$ has one fixed
point on $H$ ($w_-= \overline w_+\notin {\bf R}$) and $\Sigma$ has a 
branched point $z$ with index $q^{-1}\in {\bf N}\backslash\{0,1\}$
where $q^{-1}$ is the finite order of the stabilizer of $z$.}

\item[2.] {{\bf Parabolic.}
$|{\rm tr} \,\gamma|=2$, then $w_-=w_+\in {\bf R}$ and
the Riemann surface has a puncture. The order of the stabilizer
is now infinite, that is $q^{-1}=\infty$.}

\item[3.] {{\bf Hyperbolic.}
$|{\rm tr} \,\gamma|>2$, the fixed points are
distinct and lie on the real axis, thus $w_\pm \notin H$.  These group
elements represent handles of the Riemann surface and can be 
represented in the form $(\gamma w -w_+)/(\gamma w -w_-)=e^\lambda
(w -w_+)/(w -w_-)$, $e^\lambda\in {\bf R}\backslash\{0,1\}$.}
\end{itemize}
 
Note that if $\Gamma$ contains elliptic elements then $H/\Gamma$
is an orbifold. Furthermore,
since the parabolic points
do not belong to $H$, point $J_H(w_+)$ corresponds
to a deleted point of $\Sigma$.  
By abuse of language
we shall call both the elliptic and the 
parabolic points ramified punctures.

\vspace{0.3cm}

A Riemann surface isomorphic to the quotient $H/\Gamma$
has the Poincar\'e metric $\hat g$ as the unique metric with scalar curvature 
$R_{\hat g}=-1$ compatible with its complex structure.
This implies the uniqueness 
of the solution of the Liouville 
equation on $\Sigma$.
The Poincar\'e metric on $H$ is
\begin{equation}
d{\hat s}^2={|dw|^2\over ({\rm Im}\,w)^{2}}.
\label{tyrop}\end{equation} 
Note that $PSL(2,{\bf R})$ transformations are isometries of
$H$ endowed with the Poincar\'e metric.

An important property of $\Gamma$ is that it is
isomorphic to the fundamental group $\pi_1(\Sigma)$.
Uniformizing groups admit the following structure.
Suppose $\Gamma$ uniformizes a surface of genus $h$
with $n$ punctures and $m$ elliptic points with indices 
$2\le q_1^{-1}\le q_2^{-1}\le \ldots 
\le q_{m}^{-1}<\infty$. In this case the Fuchsian
group is generated by
$2h$ hyperbolic elements $H_1,\ldots,H_{2h}$,
$m$ elliptic elements $E_1,\ldots,E_m$
and $n$ parabolic elements $P_1,\ldots,P_n$, satisfying the relations
\begin{equation}
E_i^{q_i^{-1}}=I,\qquad
\prod_{l=1}^mE_l\prod_{k=1}^nP_k\prod_{j=1}^h
\left(H_{2j-1}H_{2j}{H_{2j-1}^{-1}}{H_{2j}^{-1}}\right)=I,
\label{iodlkjnm}\end{equation}
where the infinite cyclicity of parabolic fixed point stabilizers is
understood.

Setting $w=J_H^{-1}(z)$ in (\ref{tyrop}), 
where  $J_H^{-1}:\Sigma\to H$
is the inverse of the uniformization map, we get the Poincar\'e metric
on $\Sigma$ 
\begin{equation}
d{\hat s}^2=2{\hat g}_{z\bar z}|dz|^2=
e^{\varphi(z,\bar z)}|dz|^2,\label{pncrsk}\end{equation}
where
\begin{equation}
e^{ \varphi(z,\bar {z})}={|{J_H^{-1}(z)}'|^2\over({\rm Im}\,
J_H^{-1}(z))^2},\label{2prev}\end{equation}
which is invariant under $SL(2,{\bf R})$ fractional transformations
of $J_H^{-1}(z)$. Since
\begin{equation}
R_{\hat g}=-{\hat g}^{z\bar z}\partial_z\partial_{\bar z}
\log {\hat g}_{z\bar z},\qquad \hat g^{z\bar z}=2e^{-\varphi},
\label{sc}\end{equation}
the condition $R_{\hat g}=-1$ is equivalent 
to the Liouville equation
\begin{equation}
\partial_z\partial_{\bar {z}}\varphi(z,\bar{z})=
{1\over 2}e^{ \varphi(z,\bar{z})},\label{1}\end{equation} 
whereas the field $\widetilde \varphi=\varphi +\log \mu$, $\mu>0$,
defines a metric of constant curvature $-\mu$. Notice that the metric
$g_{z\bar z}=e^{\sigma+\varphi}/2$, with 
$\partial_z\partial_{\bar z}\sigma=0$, has scalar curvature 
$R_g=-e^{-\sigma}$. However recall that non constant harmonic functions
do not exist on compact Riemann surfaces.

\subsection{The Liouville Condition}

If $g$ is a (in general non singular) metric
on a Riemann surface of genus $h$ with $n$ parabolic
points
we have\footnote{In the following
the ${|dz\wedge d\bar z|\over 2}$ term in the surface integrals
is understood.}
\begin{equation}
\int_\Sigma \sqrt g R_g=2\pi\chi(\Sigma),
\label{gb2}\end{equation}
where
\begin{equation}
\chi(\Sigma)=2-2h-n.\label{elcrt}\end{equation}
A peculiarity of  parabolic points is that they 
do not belong to $\Sigma$, so that the singularities
in the metric and in the 
Gaussian curvature at the punctures do not appear
in $g$ and $R$ as  functions on $\Sigma$.

Let $\Sigma$ be a $n$-punctured Riemann surface
of genus $h$ and with  elliptic 
points $(z_1,\ldots,z_m)$.
Its Poincar\'e area is \cite{kra0}
\begin{equation}
\int_\Sigma \sqrt {\hat g} =2\pi\left(2h-2+n+\sum_{k=1}^m(1-q_k)\right),
\label{abelcrt}\end{equation}
where $q^{-1}_k\in {\bf N}\backslash\{0,1\}$
denotes the ramification index of $z_k$.

Let us choose the coordinates in such a way that
the metric be in the conformal form
$ds^2=2g_{z\bar z}|dz|^2$. In this case 
$g_{z\bar z}=e^\sigma \hat g_{z\bar z}$ (here we set $\gamma=1$) where 
$e^\sigma\in {\cal C}^\infty_+$  and $\hat g_{z\bar z}=e^\varphi/2$ 
is the Poincar\'e metric. Since
\begin{equation}
R_g=-2e^{-\sigma-\varphi}\partial_z\partial_{\bar z}(\varphi+\sigma)=
-e^{-\sigma}\left(1+2e^{-\varphi}\partial_z\partial_{\bar 
z}\sigma\right), \label{scrs}\end{equation}
by (\ref{gb2}) it follows that
\begin{equation}
\int_\Sigma \sqrt g R_g=-
\int_\Sigma \partial_z\partial_{\bar z}(\varphi+\sigma)=-
\int_\Sigma \partial_z\partial_{\bar z}\varphi=2\pi\chi(\Sigma).
\label{gb3}\end{equation}
Eq.(\ref{gb3}) follows from the fact that $\log f\in {\cal T}^{0,0}$
for $f\in {\cal C}^\infty_+$. This
shows that for admissible metric the contribution to $\chi(\Sigma)$
comes from the  transformation property of $\varphi$,
whereas terms such as $\int_\Sigma\sigma_{z\bar z}$, 
$e^{\sigma}\in {\cal C}^\infty_+$,  are vanishing.
A necessary condition in order
that $e^\sigma \hat g$ be an admissible metric on $\Sigma$
is that $\sigma$ satisfies the {\it Liouville condition}
\begin{equation}
\int_\Sigma\sigma_{z\bar z}=0,\label{oixaq}\end{equation}
which is a weaker condition than $e^{\sigma}\in {\cal C}^\infty_+$.
This trivial remark is useful to understand
the structure of the boundary of space of admissible metrics 
$e^{\sigma}\hat g$.
Actually it is possible to add delta-like singularities at the
scalar curvature leaving the Euler characteristic unchanged.
That is these additional singularities do not imply additional punctures on 
the surface. In particular
there are positive semidefinite $(1,1)$-differentials $g_{z\bar z}=
e^\sigma \hat g_{z\bar z}$
 that, since $\int_\Sigma \sqrt g R_g=\int_\Sigma \sqrt {\hat g} R_{\hat 
g}$, can be considered as degenerate metrics.
  An interesting case is when 
$\sigma(z)=-4\pi G(z,w)$ where $G$
is Green's function for the scalar Laplacian
with respect to the Poincar\'e metric $\hat g_{z\bar z}=e^{\varphi}/2$.
The Green
 function takes real values and has the behaviour $-{1\over 2\pi}\log|z-w|$
as $z\to w$, moreover
\begin{equation}
-\partial_z\partial_{\bar z}G(z,w)=\delta^{(2)}(z-w)-
{\sqrt {\hat g}\over 2\int_\Sigma \sqrt {\hat g}},
\label{grnfct}\end{equation}
where the $-\sqrt {\hat g}/2\int_\Sigma \sqrt {\hat g}=
e^{\varphi}/8\pi\chi$ term 
is due to the constant zero-mode of the Laplacian.
Thus, in spite of the logarithm singularity of $G$,
the contribution to $\int_\Sigma G_{z\bar z}$
coming from the delta function is cancelled
by the contribution due to the 
$e^{\varphi}/8\pi\chi$ term. Therefore 
$\int_\Sigma \partial_z\partial_{\bar z}G(z,w)=0$ and 
$\chi(\Sigma)$ is unchanged whereas
the scalar curvature becomes 
\begin{equation}
R_{\hat g}(z,\bar z)=-1\to R_g(z,\bar z)=-e^{4\pi G(z,w)}
\left(1+8\pi e^{-\varphi(z,\bar z)}\delta^{(2)}(z-w)+{1\over 
4\chi(\Sigma)}\right).\label{rdlt}\end{equation}

Note that similar remarks extend to the positive definite metric 
$e^{2\pi G(z,w)+\varphi}$. 
We conclude this digression 
by stressing that  one can modify the metric 
by adding singularities in such a way that the Euler characteristic changes. 
In this case 
one should try to define a new surface with additional punctures where
the $(1,1)$-differential is an admissible metric.

\subsection{Chiral Factorization And Polymorphicity}

By means of chiral (in general polymorphic)
functions it is possible to construct regular and non vanishing differentials
 $f^{(n,n)}\in{\cal T}^{n,n}$.
An important example is given by the expression of the 
Poincar\'e metric in terms of the inverse map $J_H^{-1}$ (which is
a chiral polymorphic scalar function) or in terms of solutions of the
uniformization equation (cfr. (\ref{dedade1})).
 Conversely, a factorized form $f^{(n,n)}=g_1(z) g_2(\bar z)$
enforces us to consider
chiral differentials whose degree is fixed by $n$ and the topology
of $\Sigma$.
It is easy to see that
the only change that $g_1$ and $g_2$ can undergo
after winding around the homology cycles of $\Sigma$
is to get a constant multiplicative factor. However
these differentials have the same degree as singlevalued
differentials, that is ${\rm deg}\,g_1={\rm deg}\,g_2=2n(h-1)$. 

Later we will see that
 similar aspects force us to consider non Abelian
monodromy for the chiral function 
arising in the construction of the Poincar\'e metric.

\subsection{$\mu$, $\chi(\Sigma)$ {And The Liouville Equation}}

Here we consider some aspects of the Liouville equation.
We start by noticing that by Gauss-Bonnet 
it follows that if $\int_\Sigma e^\varphi>0$, then the equation 
\begin{equation}
\partial_z\partial_{\bar {z}}\varphi(z,\bar{z})=
 {\mu\over 2}e^{ \varphi(z,\bar{z})},\label{dnsxtst}\end{equation}
has no solutions on surfaces with ${\rm sgn}\,\chi(\Sigma)=
{\rm sgn}\,\mu$.
In particular, on the
 Riemann sphere with $n\le 2$ punctures\footnote{The
1-punctured Riemann sphere, i.e. $\bf C$, has itself
as universal covering. For $n=2$ 
we have $J_{\bf C}:{\bf C}\to 
{\bf C}\backslash \{0\}$, $z\mapsto e^{2\pi i z}$.
Furthermore, 
${\bf C}\backslash \{0\}\cong {\bf C}/<T_1>$, where $<T_1>$ is the 
group generated by $T_1:z\mapsto z+1$.}
there are no solutions of the equation
\begin{equation}
\partial_z\partial_{\bar {z}}\varphi(z,\bar{z})=
 {1\over 2}e^{ \varphi(z,\bar{z})}.
\qquad \int_\Sigma e^\varphi>0,\label{doesnot}\end{equation}
The metric of constant curvature on $\widehat {\bf C}$
\begin{equation}
ds^2=e^{\varphi_0}|dz|^2,
\qquad e^{\varphi_0}=
{4\over \left(1+|z|^2\right)^2},\label{sphere}\end{equation}
satisfies the
 Liouville equation with the ``wrong sign'', that is
\begin{equation}
R_{\varphi_0}=1\quad\longrightarrow\quad
\partial_z\partial_{\bar {z}}\varphi_0(z,\bar{z})=
- {1\over 2}e^{ \varphi_0(z,\bar{z})}.\label{itdoes}\end{equation}
If one insists on finding a
solution of eq.(\ref{doesnot}) 
on $\widehat {\bf C}$, then  inevitably one obtains at least three
delta-singularities
\begin{equation}
\partial_z\partial_{\bar {z}}\varphi(z,\bar{z})=
 {1\over 2}e^{ \varphi(z,\bar{z})}-2\pi\sum_{k=1}^n\delta^{(2)}(z-z_k),
\qquad n\ge 3. 
\label{puoxi}\end{equation}
Since $\sigma=\log (\varphi-\varphi_0)$ does not satisfy the Liouville
condition, the $(1,1)$-differential $e^\varphi$ is not an admissible
metric on $\widehat {\bf C}$. Furthermore,
since the unique solution of the
equation $\varphi_{z\bar z}=e^\varphi/2$ on 
the Riemann sphere
is $\varphi=\varphi_0+i\pi$,
 to consider the Liouville equation on $\widehat{\bf C}$ 
gives the unphysical metric $-e^{\varphi_0}$.

This discussion shows that in order to find a solution of
eq.(\ref{doesnot}) one needs at least
three punctures, that is one must consider eq.(\ref{doesnot})
on the surface
$\Sigma=\widehat {\bf C}\backslash\{z_1,z_2,z_3\}$ where the term
$2\pi\sum_{k=1}^3\delta^{(2)}(z-z_k)$ does not appear
simply because $z_k\notin \Sigma$, $k=1,2,3$. 
In this case $\chi(\Sigma)=-1$, so that
${\rm sgn}\,\chi(\Sigma)=-{\rm sgn}\,\mu$ in agreement
with Gauss-Bonnet.

\subsection{The Inverse Map And The Uniformization Equation}

Let us now consider some aspects of the Liouville equation (\ref{1}).
As we have seen the Poincar\'e metric on $\Sigma$ is
\begin{equation}
e^{ \varphi(z,\bar {z})}={|{J_H^{-1}(z)}'|^2\over({\rm Im}\,
J_H^{-1}(z))^2},\label{2}\end{equation}
which is invariant under $SL(2,{\bf R})$ fractional transformations of 
$J^{-1}_H$. This metric is the unique solution of the Liouville 
equation.

An alternative expression for $e^\varphi$
follows by considering as universal covering of $\Sigma$ the
Poincar\'e disc $\Delta=\{z||z|<1\}$. Let us denote by 
$J_{\Delta}: \Delta \to \Sigma$ the map of uniformization.
 Since the map from $\Delta$ to $H$ is 
\begin{equation}
w=i{1-z\over z+1},\quad\qquad z\in
\Delta,\quad w\in H,\label{cgd}\end{equation}
we have
\begin{equation}
e^\varphi=4 {|{J_{\Delta}^{-1}}'|^2\over
(1-|J_{\Delta}^{-1}|^2)^2}=4|{J_{\Delta}^{-1}}'|^2\sum_{k=0}^\infty
(k+1)|{J_{\Delta}^{-1}}|^{2k}.
\label{poincdisc}\end{equation}

Both (\ref{2}) and (\ref{poincdisc}) make it evident that from the explicit
expression of the inverse map we can find the dependence of $e^\varphi$ on the 
moduli of $\Sigma$. Conversely we can express the inverse map (to within a 
$SL(2,{\bf C})$ fractional transformation) in terms of $\varphi$.
This follows from the Schwarzian equation
\begin{equation}\{J_H^{-1},z\}=
T^F(z),\qquad \label{4}\end{equation}
where
\begin{equation}
T^F(z)=\varphi_{zz}-{1\over 2}\varphi_z^2,
\label{stress1}\end{equation}
is the classical Liouville energy-momentum tensor (or Fuchsian 
projective connection) and
\begin{equation}
\{f,z\}
={f'''\over f'}-{3\over 2}\left({f''\over f'}\right)^2=
-2(f')^{1\over 2}((f')^{-{1\over 2}})'',
\label{schrz}\end{equation}
is the Schwarzian derivative. 
Note that eq.(\ref{1}) implies that 
\begin{equation}
\partial_{\bar z}T^F=0.\label{chtt}\end{equation}
In the conformal gauge the metric can be written as 
$g_{z\bar z}=e^{\varphi+\gamma\sigma}/2$, 
$e^{\gamma\sigma}\in {\cal C}^\infty_+$. In this
case the stress tensor 
\begin{equation}
T^\gamma=(\varphi+\gamma\sigma)_{zz}-
{1\over 2}(\varphi_z+\gamma\sigma_z)^2,\label{stwsigma}\end{equation}
satisfies the equation
\begin{equation}
\partial_{\bar z} T^\gamma=-e^{\varphi +\gamma\sigma}
\partial_z R_{\varphi+\gamma\sigma}.\label{nchiral}\end{equation}
Therefore $T^\gamma$ is not chiral unless 
$R_{\varphi+\gamma\sigma}$ is an antiholomorphic function.
Of course the only possibility compatible with the fact that
$e^{\gamma\sigma}\in {\cal C}^\infty_+$ is 
$R_{\varphi+\gamma\sigma}=cst$.
Another aspect of the stress tensor is
that $SL(2,{\bf C})$ transformations of
$J_H^{-1}$, while changing the Poincar\'e metric, leave
$T^F$ invariant.

On punctured surfaces there are 
non trivial global solutions of the equation $\sigma_{z\bar z}=0$, so that
in this case $\partial_{\bar z}T^\gamma=-\gamma \partial_{\bar z}
(\varphi_z\sigma_z)=-{\gamma\over 2} e^\varphi\sigma_z$. Furthermore,
since
$\partial_zR_{\varphi+\gamma\sigma}=
{\gamma\over \beta}\partial_z R_{\varphi+\beta\sigma}$,
with $\beta$ an arbitrary constant, we have
\begin{equation}
\partial_{\bar z}T^\gamma=-{\gamma\over \beta}e^{\varphi+\gamma\sigma}
\partial_z R_{\varphi+\beta\sigma}.\label{nmy}\end{equation}

Let us define the {\it covariant Schwarzian operator}
\begin{equation}
{\cal S}^{(2)}_f=2(f')^{1\over 2}\partial_z (f')^{-1}
\partial_z(f')^{1\over 2},\label{sf1}\end{equation}
mapping $-1/2$- to $3/2$-differentials.
Since
\begin{equation}
{\cal S}^{(2)}_f\cdot\psi=\left(2\partial^2_z+\{f,z\}\right)\psi,
\label{proies}\end{equation}
the Schwarzian derivative can be written as
\begin{equation}
\{f,z\}={\cal S}^{(2)}_f\cdot 1.\label{eqar}\end{equation}
The operator ${\cal S}^{(2)}_f$ is invariant under  $SL(2,{\bf C})$
fractional transformations of $f$, that is
\begin{equation}
{\cal S}^{(2)}_{\gamma \cdot f}={\cal S}^{(2)}_f,\qquad \gamma\in SL(2,{\bf C}).
\label{smmtr4}\end{equation}
Therefore, if the transition functions of $\Sigma$ are 
linear fractional transformations, then $\{f,z\}$ 
transforms as a quadratic differential.
However, except in the case of projective coordinates,
the Schwarzian derivative does not transform covariantly 
on $\Sigma$. This is evident by (\ref{eqar}) since
in flat spaces only (e.g. the torus)
a constant can be considered as a $-1/2$-differential.

Let us consider the equation 
\begin{equation}
{\cal S}^{(2)}_f\cdot\psi=0.\label{sclin}\end{equation}
To find two independent solutions we set 
\begin{equation}
(f')^{1\over 2}\partial_z (f')^{-1}
\partial_z(f')^{1\over 2}\psi_1=(f')^{1\over 2}\partial_z (f')^{-1}
\partial_z cst=0,\label{c1}\end{equation}
 and
\begin{equation}
(f')^{1\over 2}\partial_z (f')^{-1}
\partial_z(f')^{1\over 2}\psi_2=(f')^{1\over 2}\partial_z  cst=0,
\label{c2}\end{equation}
so that the solutions of (\ref{sclin}) are 
\begin{equation}
\psi_1=cst\,(f')^{-{1\over 2}},\quad \psi_2=cst\, 
f(f')^{-{1\over 2}}.\label{hel}\end{equation}
Since $\psi_2/\psi_1=cst\,f$,
to find the solution of the Schwarzian equation
$\{f,z\}=g$ is equivalent
to solve the linear equation
\begin{equation}
\left(2\partial^2_z+g(z)\right)\psi=0.
\label{ficxk}\end{equation}
We stress that the ``constants'' in the linear 
combination $\phi=a\psi_1+b\psi_2$ 
admit a $\bar z$-dependence provided that
$\partial_z a=\partial_z b=0$.

The inverse map is locally univalent,
that if $z_1\ne z_2$ then ${J_H^{-1}}(z_1)\ne
{J_H^{-1}}(z_2)$.
A related characteristic of $J_H^{-1}$ is
that under a winding of $z$ around 
non trivial cycles of $\Sigma$ the point $J_H^{-1}(z)\in H$ moves from a 
representative ${\cal D}$ of the fundamental domain  to an equivalent point 
of another representative\footnote{This property of $J_H^{-1}$ makes evident
its univalence as function on $\Sigma$.} ${\cal D}'$. On the other hand, since 
$\Gamma$ is the automorphism group of $J_H$, it follows that after winding 
around non trivial cycles of $\Sigma$ the 
inverse map transforms in the linear fractional way 
\begin{equation}
J_H^{-1}\longrightarrow \gamma\cdot J_H^{-1}=
{a J_H^{-1} +b\over c J_H^{-1}+d},\quad
\left(\begin{array}{c}a\\c
\end{array}\begin{array}{cc}b\\d\end{array}\right)\in \Gamma.
\label{Jtr}\end{equation}
However note that (\ref{smmtr4}) 
guarantees that, in spite of the polymorphicity 
(\ref{Jtr}), the classical Liouville stress tensor 
$T^F={\cal S}^{(2)}_{J^{-1}_H}\cdot 1$ is singlevalued.

As we have seen 
one of the important properties of the Schwarzian derivative is that
the Schwarzian equation (\ref{4}) can be linearized. 
Thus if $\psi_1$ and $\psi_2$ are 
linearly independent solutions of the  {\it uniformization equation}
\begin{equation}
\left({\partial^2\over \partial z^2}+{1\over
2}T^F(z)\right)\psi(z)=0,
\label{new1}\end{equation}
then $\psi_2/\psi_1$ is a solution of eq.(\ref{4}). That is,
up to a $SL(2,{\bf C})$
linear fractional transformation, we have
\begin{equation}
J_H^{-1}=\psi_2/\psi_1.
\label{djhq}\end{equation}
Indeed by (\ref{c1},\ref{c2}) 
it follows that 
\begin{equation}
\psi_1={({J_H^{-1}}')}^{-{1\over 2}},\qquad \psi_2=
{({J_H^{-1}}')}^{-{1\over 2}} J_H^{-1}, \label{solst}\end{equation}
are independent solutions of (\ref{new1}).
Another  way to prove (\ref{djhq}) 
is to write eq.(\ref{new1}) in the equivalent form
\begin{equation}
      (f')^{1\over 2}\partial_z (f')^{-1}
\partial_z(f')^{1\over 2}\psi=0,
\qquad f\equiv J_H^{-1},\label{qvlnt}\end{equation}
and then to set $z=J_H(w)$. In this case
(\ref{new1}) becomes the trivial equation on $H$
\begin{equation}
{w'}^{3/2}\partial^2_w\phi=0.
\label{qvlnt1}\end{equation}
For any choice of the two linearly independent
solutions we have $\phi_2/\phi_1=w$
up to an $SL(2,{\bf C})$ transformation.
Going back to $\Sigma$ we get
$J_H^{-1}=\psi_2/\psi_1$.

Note that any $SL(2,{\bf R})$ transformation 
\begin{equation}
\left(\begin{array}{c} \psi_1\\ \psi_2 \end{array}\right)\longrightarrow
\left(\begin{array}{c} \widetilde\psi_1\\ \widetilde\psi_2 \end{array}\right)=
\left(\begin{array}{c}a\\c
\end{array}\begin{array}{cc}b\\d\end{array}\right)
\left(\begin{array}{c} \psi_1\\ \psi_2 \end{array}\right),
\label{transfofpsi}\end{equation}
induces a linear fractional transformation of $J^{-1}_H$.
Therefore the invariance of $e^\varphi$ under $SL(2,{\bf R})$ 
linear fractional transformations of $J^{-1}_H$ corresponds to its
invariance for $SL(2,{\bf R})$ linear transformations 
of\footnote{Note that the Poincar\'e metric is 
invariant under $SL(2,{\bf R})$ fractional transformations
of  $J_H^{-1}$ whereas the Schwarzian derivative 
$T^F(z)=\{J_H^{-1},z\}$ is invariant for
$SL(2,{\bf C})$ transformations of $J_H^{-1}$.
Thus the identification $J_H^{-1}=\psi_2/\psi_1$ is up to
a $SL(2,{\bf C})$ transformation.} $\psi_1,\psi_2$. 
This leads us to express $e^{-k\varphi}$ as 
\begin{equation}
e^{-k\varphi}=(-4)^{-k}\left(\overline\psi_1\psi_2-
\overline\psi_2\psi_1\right)^{2k},
\label{dedade1}\end{equation}
in particular, when $2k$ is a non negative integer, we get
\begin{equation}
e^{-k\varphi}=
4^{-k}\sum_{j=-k}^k (-1)^{j} C_{2k}^{j+k} \overline\psi_1^{k+j}
\psi_1^{k-j}\psi_2^{k+j}\overline\psi_2^{k-j},
\quad 2k\in{\bf Z}^+,\qquad
C_k^j={k!\over j!(k-j)!}.\label{dedade01}\end{equation}
On the other hand, since we can choose 
$\psi_2=\psi_1\int \psi_1^{-2}$, we have
\begin{equation}
e^{-k\varphi}=(-4)^{-k}|\psi|^{4k}
\left(\int\psi^{-2}-\int\overline\psi^{-2}\right)^{2k},\qquad \forall k,
\label{usty}\end{equation}
with
\begin{equation}
\psi=a\psi_1(1+b\int\psi_1^{-2}), \qquad
a\in{\bf R}\backslash \{0\},\quad b\in {\bf R}.\label{piel}\end{equation}
We note that the ambiguity in the definition of $\int^z\psi^{-2}$ 
reflects the polymorphicity of $J_H^{-1}$.
This property of $J_H^{-1}$ implies that, under a
winding around non trivial loops, a solution of
(\ref{new1}) transforms in a linear combination involving
itself and another (independent) solution.

It is easy to check that 
\begin{equation}
\left(\partial^2_z+{1\over 2} 
T^F(z)\right)e^{-\varphi/2}=0,
\label{nullvectors}\end{equation}
which shows that the uniformization equation
has the interesting property of admitting
singlevalued solutions.
The reason is that the $\bar z$-dependence of
$e^{-\varphi/2}$ arises through the coefficients 
$\overline\psi_1$ and $\overline\psi_2$ in the linear combination
of $\psi_1$ and $\psi_2$.

Since
$[\partial_{\bar z},{\cal S}^{(2)}_{J_H^{-1}}]=0$, the
singlevalued solutions of the uniformization equation are
\begin{equation}
\left(\partial^2_z+{1\over 2} 
T^F(z)\right)\partial_{\bar z}^le^{-\varphi/2}=0, 
\qquad l=0,1,\ldots .
\label{nullvectors03}\end{equation}
Thus, since $e^{-\varphi}$ and $e^{-\varphi}\varphi_{\bar z}$
are linearly independent solutions of eq.(\ref{new1}),
their ratio solve the Schwarzian equation
\begin{equation}
\{\varphi_{\bar z},z\}=T^F(z).\label{oiqp}\end{equation}
Higher order derivatives $\partial^l_{\bar z}e^{-\varphi/2}$,
$l\ge 2$, are linear combinations of
$e^{-\varphi/2}$ and $e^{-\varphi/2}\varphi_{\bar z}$
with coefficients depending on $\overline T^F$ and its derivatives;
for example 
\begin{equation}
\partial^2_{\bar z}e^{-\varphi/2}=-{\overline T^F\over 2}
e^{-\varphi/2}.\label{frxmpl}\end{equation}
In particular if $\psi_2(z)=\overline T^F\psi_1(z)$
then, in spite of the fact that $\overline T^F$ is not a constant
on $\Sigma$, $\psi_1$ and $\psi_2$ are linearly dependent solutions of
eq.(\ref{new1}). A check of the linear dependence of
$\psi_1$ from $\psi_2$ follows from the fact that
\begin{equation}
\{\psi_2/\psi_1,z\}=
\{\overline T^F,z\}=0\ne T^F(z).\label{netot}\end{equation}

Let us show what happens if
one sets $J_H^{-1}=\psi_1/\psi_2$
without considering the remark made in the previous footnote.
As solutions of the uniformization equation,
we can consider $\psi_1=e^{-\varphi/2}$ and an arbitrary
solution $\psi_2$ such that 
$\partial_z \left(\psi_2/\psi_1\right)=0$.
Since
$\partial_{\bar z}\left(e^{-\varphi/2}/\psi_2\right)\ne 0$,
in spite of the fact that
$\{e^{-\varphi/2}/\psi_2,z\}=T^F$,
we have $J_H^{-1}\ne \psi_1/\psi_2$.

We conclude the analysis of the uniformization
equation by summarizing some useful expressions 
for the Liouville stress tensor
$$
{T^F}=\left\{J_H^{-1},z\right\}= \{\varphi_{\bar z},z\}=
2\left({J_H^{-1}}'\right)^{1\over 2} 
\partial_z{1\over {J_H^{-1}}'}\partial_z \left({J_H^{-1}}'\right)^{1\over 2}
\cdot 1=
2e^{\varphi/2}
\partial_z
e^{-\varphi}
\partial_z
e^{\varphi/2}\cdot 1 $$
\begin{equation}
=2{\left({e^{-\varphi/2}\over\psi_2}\right)'}^{1\over2}
\partial_z 
{\left({e^{-\varphi/2}\over\psi_2}\right)'}^{-1}
\partial_z
{\left({e^{-\varphi/2}\over\psi_2}\right)'}^{1\over2}
\cdot 1=-2e^{\varphi/2}\left( e^{-\varphi/2}\right)''
=-2 \psi^{-1}\psi'',\label{four}\end{equation}
with $\psi$ given in (\ref{piel}) and $\psi_2$ an arbitrary 
solution of eq.(\ref{new1}) such that 
$\partial_z\left(e^{-\varphi/2}/\psi_2\right)\ne 0$.

\subsection{Remarks On 
$e^{\varphi_A}= {|{A}'|^2\over \left({\rm Im}\,A\right)^2}$}

Sometimes in current literature it is stated that
the solution of the Liouville equation is 
\begin{equation}
e^{\varphi_A}= {|{A}'|^2\over \left({\rm Im}\,A\right)^2},
\label{AAA}\end{equation}
with $A$ a generic holomorphic function. 
However the uniqueness of the solution of the Liouville equation 
implies that $A(z)$ 
is the inverse map of the uniformization
which is unique up to $SL(2,{\bf R})$ fractional transformations.
Let us show what happens
if $A$ is considered to be an arbitrary well-defined chiral function
on a compact Riemann surface.
First of all according to the Weierstrass gap theorem
a meromorphic function $f^{(0)}$, with 
divisor in general position,
 has at least $h+1$ zeroes\footnote{The restriction 
to ``points in general position'' means that we are  not
considering Weierstrass points in ${\rm Div}\, f^{(0)}$.}.
Thus, since ${\rm deg}\,f^{(0)}=0$,
$f^{(0)}$ has at least $h+1$ poles.
Since $\partial_{\bar z} z^{-1}\sim\pi\delta^{(2)}(z)$,
it follows that if $A(z)$ were  a well-defined $0$-differential 
then it would induce singularities in the
scalar curvature at the divisor of
$A$, so that $R_{\varphi_A}\ne -1$.
Furthermore $e^{\varphi_A}$ itself is singular when
${\rm Im}\, A(z)=0$ 
and will degenerate for the zeroes of $A'(z)$. Therefore,
in order that $\varphi_A$ be the solution of the Liouville equation 
\begin{equation}
R_{\varphi_A}=-1,\label{scgag}\end{equation}
the field $A(z)$ must be a chiral and linearly polymorphic
function. In particular, under the action of the fundamental 
group $\pi_1(\Sigma)$, $A(z)$ must transform in a linear fractional
way with the coefficients of the transformation in the Fuchsian group
$\Gamma$ whose elements are fixed by the moduli of $\Sigma$. 

Notice that $A(z)$ cannot simply be a holomorphic
nowhere vanishing function with constant multivaluedness. 
In this case the monodromy is Abelian. From the analytic point of view
the commutativity of group monodromy has the effect
of giving a metric with singularities. This, for 
example, follows from the fact that if $A\to cst\,A$, then
$A'/A$ would be a well-defined one-differential so that
\begin{equation}
\# zeroes \,(A'/A)=\# poles\,(A'/A) +2(h-1)>0.\label{index09}
\end{equation}
On the other hand since $A$ must be a holomorphic nonvanishing
function it follows that ${\rm Div} A'=
{\rm Div }(A'/A)$. Thus $e^{\varphi_A}$ would be
a degenerate metric since,
if ${\rm Im}\,A(P)\ne 0$, 
$\forall P\in {\rm Div} A'$, then
\begin{equation}
\# zeroes \,\left(e^{\varphi_A}\right)=4(h-1)>0.
\label{isnotpoinc}\end{equation}

Unfortunately no one has succeeded in writing down the explicit
form of the inverse map in terms of the moduli of $\Sigma$.
This is the uniformization problem. In section 5 we will introduce
a new set of Fourier modes in higher genus whose properties suggest that
they are strictly related to the underlying Fuchsian group.

\subsection{On The Standard Approach To Liouville Gravity}

We conclude this section by considering some aspects concerning
the standard approach to Liouville gravity. Let us begin
by noticing that if one parametrizes the metric
in the form $g=e^{\gamma\sigma} \hat g$, where 
$\hat g$ is a background metric (in particular $\hat g$
is a positive definite $(1,1)$-differential),
then  $e^{\gamma\sigma}\in {\cal C}^\infty_+$.
The Liouville action in harmonic coordinates reads
\begin{equation}
S=\int_{\Sigma}\left(
|\partial_z \sigma|^2+{1\over \gamma}\sqrt{\hat g}R_{\hat g}
\sigma+{\mu\over 2\gamma^2}\sqrt{\hat g}e^{\gamma\sigma}\right).
\label{wrtbcm3}\end{equation}
Sometimes it is stated that at the classical level 
$\gamma\sigma$ transforms as in (\ref{tr6})
whereas after quantization the logarithm term 
is multiplied by a constant related to the central charge. Actually, 
to perform the surface integration
$\hat g$ must be a $(1,1)$-differential
and $e^{\gamma\sigma}\in {\cal T}^{0,0}$.
Therefore
it is unclear what the meaning
of $S$ is if one considers $e^{\gamma\sigma}\in{\cal T}^{1,1}$.
Another standard choice is to set $d{\hat s}^2=cst|dz|^2$ on a patch. 
Once again, since it is not possible to make this choice on the whole 
manifold, with this prescription the surface integral is undefined.
A way to (partly) solve these problems
is to set (formally) $\hat g_{z\bar z}=1$ 
and then consider 
$ds^2=e^{\gamma\sigma}|dz|^2$,
so that $e^{\gamma\sigma}\in{\cal T}^{1,1}$. In this case
$R_{\hat g}$ is formally zero and
the integrand in (\ref{wrtbcm3}) reduces to
\begin{equation}
F=|\partial_z \sigma|^2+{\mu\over 2\gamma^2}e^{\gamma\sigma}.
\label{bhooh}\end{equation}
The transition from the integrand in (\ref{wrtbcm3}) to $F$ is implicitly
assumed by some authors (see for example equations (1.2) and (2.1)
in the interesting paper \cite{srg}). 
However, since $|\partial_z \sigma|^2$ does not transform
covariantly, one must add ``boundary terms'' to $\int_\Sigma F$
in order to get a well-defined action.
This has been done in  \cite{0}. In the case of surfaces with 
punctures this boundary term corresponds to a regularization term
which is crucial in fixing the scaling properties of $S$. This 
regularization procedure is related to the fact that 
negatively curved surfaces
(the realm of Liouville theory)
have both ultraviolet and infrared cutoffing 
properties. The coupling between cutoff, regularization terms in the 
Liouville action, modular anomaly is a highly non trivial (and 
interesting) aspect which is related to the properties
of univalent functions
(e.g. Koebe 1/4-theorem)
that we will discuss in the last section in the context of 2D quantum
gravity. We notice that
this subject is related to classical and quantum chaos on Riemann 
surfaces.

In the operator formulation of
CFT one considers the solutions of the classical equation of motion with
allowed singularities at the points where the $in$ and $out$ vacua
are placed. This allows one to consider
 non trivial solutions of the equation $\phi_{z\bar z}=0$.
In Liouville theory it is not possible to compute the asymptotics of the
stress tensor by standard CFT techniques. 
The reason is that the OPE in CFT is based on free fields techniques
where $<X(z)X(w)>\sim -\log (z-w)$. This explains
why it is difficult to recognize what the vacuum of Liouville theory is.

The known results in quantum Liouville theory essentially concern the 
formulation on the sphere and the torus. To get insight about 
the continuum formulation of 2D gravity in higher genus is an outstanding 
problem. To understand the difficulties that one meets with respect to the 
$h=0,1$ cases we summarize few basic facts.

\begin{itemize}

\item[$h\le 1$] {A feature of $\widehat {\bf C}$ with respect to higher 
genus surfaces is that its universal covering is $\widehat {\bf C}$ itself. 
Therefore metrics on 
$\widehat {\bf C}$ and on its universal covering coincide.
This partly explains why in this case
 computations are easier to be done.
In the torus case  ``Liouville theory'' is free, the reason is that
the metric of constant curvature $e^\varphi$ satisfies the
equation $\varphi_{z\bar z}=0$, that is $\varphi=cst$.
Therefore to quantize 2D gravity on the torus it is sufficient
to use standard CFT techniques and to impose positivity on the Liouville
field. This can easily be done because the Fourier modes on the torus 
are explicitly known\footnote{In section 5 we 
will consider the problem of formulating higher genus Fourier analysis.}.}

\item[$h\ge 2$] {In the higher genus case the metric of constant curvature
on $\Sigma$ has a richer geometrical structure with respect to
the Poincar\'e metric on its universal covering $H$. Thus in quantizing
the theory we must consider the geometry of the
moduli space or, which is the same, the geometry
of Fuchsian groups (which is encoded in $J_H^{-1}$). 
To get results in a way similar to
those derived on $\widehat {\bf C}$ we have to shift our
attention from $\Sigma$ to the upper half plane whose 
Poincar\'e metric is explicitly known (see (\ref{tyrop})).
In this case we are not considering the
underlying topology and geometry of the terms in the genus
expansion.
Nevertheless 
non perturbative results enjoy similar properties.
This similarity suggests to formulate a 
non perturbative approach to 2D gravity
based on a sort of path-integral
formulation of 2D gravity on $H$.
 The reason for this is that both $H$ and the Poincar\'e metric
on it are universal (non perturbative) 
objects underlying the full genus expansion.}

\end{itemize}

\mysection{Covariant Holomorphic Operators, 
Classical Liouville Action And Normal Ordering}

Here we introduce a set of operators ${\cal S}^{(2k+1)}_{J_H^{-1}}$
corresponding to $\partial_z^{(2k+1)}$ 
covariantized by means of $J_H^{-1}$. We stress that univalence
of $J_H^{-1}$ implies that these operators are holomorphic.
Next, we derive the chiral (polymorphic) and non chiral eigenfunctions
for a set of operators related to ${\cal S}^{(2k+1)}_{J_H^{-1}}$.
An interesting property of these operators is that
the (generalized) harmonic Beltrami differentials 
$\mu_{harm}^{(2k+1)}$ (see eq.(\ref{gfhdte}))
are in their kernel
\begin{equation}
{\cal S}_{J_H^{-1}}^{(2k+1)}\mu_{harm}^{(2k+1)}=0.
\label{gfhdteaaa}\end{equation}

We consider the cocycles associated to 
${\cal S}^{(2k+1)}_{J_H^{-1}}$. 
In this framework we show that the normal ordering for 
operators defined on $\Sigma$ is related to {\it classical} Liouville theory.
The univalence of $J_H^{-1}$ allows us
to get time-independence and locality for the cocycles.

The holomorphicity of the covariantization above
allows us to show that the chirally split anomaly of CFT reduces
to the Krichever-Novikov ($KN$) cocycle. This suggests to introduce 
higher order anomalies given as surfaces integrals of
$(1,1)$-forms defined in terms of ${\cal S}^{(2k+1)}_{J_H^{-1}}$.
These anomalies are related to the uniformization theory of 
vector bundles on Riemann surfaces (${W}$-algebras).
Remarkably, eq.(\ref{gfhdteaaa}) implies 
that these anomalies (including the
standard chirally split anomaly) vanish in the case one considers 
deformation of the complex structure induced by (generalized)
harmonic Beltrami differentials.

\subsection{Higher Order Schwarzian Operators}

Let us start by noticing that since
\begin{equation}
e^{-k\varphi}=
|{J_H^{-1}}'|^{-2k}\left({J_H^{-1}-\overline{J_H^{-1}}\over 2i}
\right)^{2k},\label{minusk}\end{equation}
it follows that the negative powers of the Poincar\'e metric satisfy the
higher order generalization of eq.(\ref{nullvectors})
\begin{equation}
{\cal S}^{(2k+1)}_{J_H^{-1}}\cdot e^{-k\varphi} =0,
\qquad k=0,{1\over 2}, 1,\ldots,
\label{covuniform2}\end{equation}
with ${\cal S}^{(2k+1)}_f$ the higher order covariant Schwarzian operator
\begin{equation}
{\cal S}^{(2k+1)}_f=(2k+1)
(f')^k \partial_z (f')^{-1}\partial_z (f')^{-1}\ldots
\partial_z (f')^{-1}\partial_z (f')^k,\label{cvprtr}\end{equation}
where the number of derivatives is $2k+1$. 
We stress that univalence of ${J_H^{-1}}$ implies holomorphicity 
of the ${\cal S}^{(2k+1)}_{J_H^{-1}}$ operators.
Eq.(\ref{covuniform2}) 
is manifestly covariant and singlevalued on 
$\Sigma$. Furthermore it can be proved that 
the dependence of
${\cal S}^{(2k+1)}_f$ on $f$ appears only through 
${\cal S}^{(2)}_f\cdot 1=\{f,z\}$ and its derivatives; for example
\begin{equation}
{\cal S}^{(3)}_{J_H^{-1}}=
3\left(\partial_z^3+2T^F\partial_z+{T^F}'\right),
\label{covuniform16}\end{equation}
which is the second symplectic structure of the KdV equation.
A nice property of the equation 
${\cal S}^{(2k+1)}_{J_H^{-1}}\cdot \widetilde\psi=0$
is that 
its projection on $H$ is the trivial equation
\begin{equation}
(2k+1){w'}^{k+1}\partial_w^{2k+1}\psi=0,\qquad w\in H,
\label{jhdlkmnhg}\end{equation}
where $w=J_H^{-1}(z)$.
This makes evident why only for $k>0$ 
it is possible to have finite expansions of $e^{-k\varphi}$  such as
in eq.(\ref{dedade01}).
The reason is that the solutions of eq.(\ref{jhdlkmnhg}) are 
$\{w^j|j=0,\ldots ,2k\}$ so that
the best thing we can do is to consider linear combinations 
of positive powers of the non chiral solution ${\rm Im}\, w$ which is 
just the square root of inverse of the Poincar\'e metric on $H$.

The $SL(2,{\bf C})$ invariance of the Schwarzian derivative implies that
\begin{equation}
{\cal S}^{(2k+1)}_{J_H^{-1}}={\cal S}^{(2k+1)}_{\varphi_{\bar z}},
\qquad k=0,{1\over 2}, 1,\ldots .
\label{covuniform20}\end{equation}
On the other hand by Liouville equation
\begin{equation}
{\cal S}^{(2k+1)}_{J_H^{-1}}={\cal S}^{(2k+1)}_{\varphi_{\bar z}}
=(2k+1)e^{k\varphi}\partial_ze^{-\varphi}\partial_ze^{-\varphi}
\ldots \partial_ze^{-\varphi}\partial_ze^{k\varphi},
\qquad k=0,{1\over 2}, 1,\ldots .
\label{covuniforam20}\end{equation}
In the following we will use this property of ${\cal S}^{(2k+1)}_{J_H^{-1}}$ 
to construct the eigenfunctions for the operator
\begin{equation}
{\cal Q}^{(2k+1)}_{\varphi_{\bar z}}=
{\cal S}^{(2k+1)}_{\varphi_{\bar z}}
\left(2\varphi_{\bar z}e^{-\varphi}\right)^{2k+1},
\label{ocihwe3}\end{equation}
and for its chiral analogous
\begin{equation}
{\cal Q}^{ch(2k+1)}_{J_H^{-1}}=
{\cal S}^{(2k+1)}_{J_H^{-1}}
\left(\partial_z \log J_H^{-1}\right)^{-2k-1}.\label{lksl}\end{equation}

\subsection{Eigenfunctions Of ${\cal Q}^{(2k+1)}_{\varphi_{\bar z}}$
And ${\cal Q}^{ch(2k+1)}_{J_H^{-1}}$}

Since
\begin{equation}
[\partial_{\bar z}, {\cal S}^{(2k+1)}_{J_H^{-1}}]=0,
\label{chgl}\end{equation}
it follows that besides $e^{-k\varphi}$
other singlevalued solutions of 
${\cal S}^{(2k+1)}_{J_H^{-1}}\cdot \psi=0$ have
the form $\partial^l_{\bar z} e^{-k\varphi}$.
However notice that by (\ref{covuniform20}) it follows that
the set of singlevalued differentials
\begin{equation}
\psi_l=\left(2\varphi_{\bar z}\right)^l e^{-k\varphi},
\qquad l=0,\ldots,2k,\label{ocihwe}\end{equation}
is a basis of solutions of
${\cal S}^{(2k+1)}_{J_H^{-1}}\cdot \psi=0$.
To see this it is sufficient to substitute $\psi_l$ in the RHS
of (\ref{covuniforam20}) and systematically use the Liouville equation
\begin{equation}
e^{-\varphi}\partial_z(2\varphi_{\bar z})^l=l(2\varphi_{\bar z})^{l-1}.
\label{ofdhwd}\end{equation}
In the intersection of two patches
$(U,z)$ and $(V,w)$ the field
$\psi_l$ transforms as
\begin{equation}
\left(2\varphi_{\bar z}(z,\bar z)\right)^l e^{-k\varphi(z,\bar z)}=
\left(2\widetilde\varphi_{\bar w}(w,\bar w)+
2\bar w_{\bar z \bar z}/(\bar w_{\bar z})^2\right)^l 
e^{-k\widetilde\varphi(w,\bar w)}w_z^{-k}\bar w_{\bar z}^{-k+l},
\label{colk}\end{equation}
that is $\psi_l$ decomposes into a sum of solutions for the
covariant operator ${\cal S}^{(2k+1)}_{J_H^{-1}}$ written
in the patch $V$. 

Let us consider the chiral (i.e. such that $\partial_{\bar 
z}\phi^{(-k)}=0$) solutions of
\begin{equation}
{\cal S}_{J_H^{-1}}^{(2k+1)}\cdot \phi^{(-k)}=0.
\label{podjdj}\end{equation}
We have $\phi^{(-k)}_l=(J_H^{-1})^l({J_H^{-1}}')^{-k}$
$l=0,\ldots , 2k$.
Note that $\phi^{(-k)}_l$ and $\psi_l$ have the common property of 
generating other solutions either by changing patch (eq.(\ref{colk}) for 
$\psi_l$) or, in the case of $\phi^{(-k)}_l$, by running around cycles.
For example, in the case $k=1/2$, starting from 
$\phi^{(-1/2)}_1$, to get the other linearly independent solution
$\phi^{(-1/2)}_2$ it is sufficient to perform a non trivial winding 
around $\Sigma$.

Since
\begin{equation}
{\cal S}^{(2k+1)}_{J_H^{-1}}\cdot 
\left(2\varphi_{\bar z}\right)^{2k+l} 
e^{-k\varphi}=\lambda_l
\left(2\varphi_{\bar z}\right)^{l-1}
e^{(k+1)\varphi},
\qquad l\in {\bf Z},\label{ocihwe1}\end{equation}
where
\begin{equation}
\lambda_l={(2k+1)(2k+l)(2k+l-1)\ldots(l+1)l},\qquad l\in{\bf Z},
\label{uql}\end{equation}
it follows  that the singlevalued differentials
\begin{equation}
\psi_l=(2\varphi_{\bar z})^{l-1}e^{(k+1)\varphi},\qquad l\in {\bf Z},
\label{pdod}\end{equation}
are eigenfunctions of ${\cal Q}^{(2k+1)}_{\varphi_{\bar z}}$
\begin{equation}
{\cal Q}^{(2k+1)}_{\varphi_{\bar z}}\cdot\psi_l=\lambda_l \psi_l,
\qquad l\in{\bf Z}.
\label{kjx}\end{equation}
Note that $\psi_{-2k},\ldots,\psi_0$ are the zero modes of 
${\cal Q}^{(2k+1)}_{\varphi_{\bar z}}$. Furthermore,
 eq.(\ref{kjx}) is invariant under
the substitution, $\psi_l\to {\cal F}\psi_l$, where
${\cal F}$ is an arbitrary solution of
\begin{equation}
\partial_z{\cal F}=0.\label{oihp}\end{equation}
Since the Liouville stress tensor satisfies the equations
$\partial_z \partial_{\bar z}^n {\overline T^F}=0$, $ n=0,1,2,\ldots$,
the general solution of (\ref{oihp})
depends on $\overline T^F$ and its derivatives. However,
taking into account polymorphic differentials,
the general solution of eq.(\ref{oihp}) has the form
\begin{equation}
{\cal F}\equiv {\cal F}\left(\overline\psi_k,\overline\psi_k',
\overline\psi_k'',\ldots\right),\quad k=1,2,
\qquad '\equiv \partial_{\bar z},
\label{hnl}\end{equation}
where $\overline\psi_1$ and $\overline\psi_2$ are two solutions of
\begin{equation}
\left(\partial^2_{\bar z}+{1\over 2} 
\overline T^F\right)\overline\psi=0,
\label{nullvectors3}\end{equation}
such that $\partial_z\overline\psi_1=
\partial_z\overline\psi_2=0$ and 
$\partial_{\bar z}\left(\overline\psi_1/\overline\psi_2\right)\ne 0$.

The differentials
\begin{equation}
\psi_l^{ch}=\left(J_H^{-1}\right)^{l-1}\left({J_H^{-1}}'\right)^{k+1},
\qquad l\in{\bf Z},
\label{hd}\end{equation}
are the chiral analogous of $\psi_l$.
Indeed they 
satisfy the equation
\begin{equation}
{\cal Q}^{ch(2k+1)}_{J_H^{-1}}\cdot\psi_l^{ch}=\lambda_l \psi_l^{ch},
\qquad l\in{\bf Z},
\label{kjxb}\end{equation}
where $\lambda_l$ is given in (\ref{uql}).
A property of ${\cal S}_{J_H^{-1}}^{(2k+1)}$ is
\begin{equation}
\left[{\cal S}_{J_H^{-1}}^{(2k+1)},\overline\psi^{(n)}\right]=0,
\label{udxdq}\end{equation}
with $\psi^{(n)}$ a holomorphic $n$-differential. This implies that the 
(generalized) harmonic Beltrami differentials satisfy the equation
\begin{equation}
{\cal S}_{J_H^{-1}}^{(2k+1)}\mu_{harm}^{(2k+1)}=0,\qquad
\mu_{harm}^{(2k+1)}=\overline\psi^{(2k+1)}e^{-k\varphi}.
\label{gfhdte}\end{equation}
This equation will be useful in recovering the kernel of the chirally
split anomaly arising in CFT. Note that $\mu_{harm}^{(3)}$ is a
standard harmonic Beltrami differential.

The operator ${\cal S}^{(3)}_{f_{BA}}$, where
\begin{equation}
\partial_z f_{BA}= e^{-1}_{BA},\label{baha2}\end{equation}
with $e_{BA}$  a {\it Baker-Akhiezer vector field}, appears in the formulation
of the covariant KdV in higher genus \cite{npb,monte}.
The non holomorphic 
operators ${\cal S}^{(2k+1)}_{f_{BA}}$,  $k\in {\bf Z}_+$ define 
cocycles on Riemann surfaces \cite{npb}.
Similar operators, not related to uniformization, have been considered
also in \cite{DIZ} (see also \cite{GI}).

\subsection{Normal Ordering On $\Sigma$ And Classical Liouville Action}

Let us now consider
the meromorphic $n$-differentials
$\psi_j^{(n)}$ proposed in \cite{kn}. 
They are the higher genus analogous 
of the Laurent monomials $z_+^{j-n}$.
By means of $\psi_j^{(n)}$ we can expand
holomorphic differentials on $\Sigma\backslash \{P_+,P_-\}$. 
Their relevance for an operator approach which mimics
the radial quantization on the Riemann sphere has been 
shown in \cite{cmp}.

In terms of local coordinates
$z_\pm$ vanishing at $P_\pm\in \Sigma$ 
the basis reads 
\begin{equation}
\psi_j^{(n)}(z_\pm)(dz_\pm)^n=a_j^{(n)\pm}
z_\pm^{\pm j -s(n)}\left(1+{\cal 
O}(z_\pm)\right)\left(dz_\pm\right)^n, 
\quad s(n)={h\over 2}-n(h-1),\label{onehalf}\end{equation}
where $j\in{\bf Z}+h/2$ and $n\in{\bf Z}$. 
The $(dz_\pm)^n$ term has been included
to emphasize that $\psi_j^{(n)}(z)$ transforms as
$\partial_z^n$.
By the Riemann-Roch theorem $\psi_j^{(n)}$ is uniquely determined
by fixing one of the constants $a_j^{(n)\pm}$ (to choose the value of
$a_j^{(n)+}$ fixes $a_j^{(n)-}$ and vice versa). In the 
following we set $a_j^{(n)+}=1$.
There are few exceptions to (\ref{onehalf}) concerning essentially
the $h=1$ and $n=0,1$ cases \cite{kn}.
The  expression of this basis in terms of theta functions reads \cite{cmp}
\begin{equation}
\psi_j^{(n)}(z)=C_j^{(n)}
\Theta\left(I(z)+{\cal D}^{j;n}|\Omega\right)
 {\sigma(z)^{2n-1} E(z, P_+)^{j-s(n)}\over
 E(z,P_-)^{j+s(n)}}, 
\label{psij}\end{equation}
where 
\begin{equation}
{\cal D}^{j;n}=\left(j-s(n)\right)I(P_+)-
\left(j+s(n)\right)I(P_-)+(1-2n)\Delta,
\label{polnmh}\end{equation}
and the constant $C_j^{(n)}$ is fixed by  
the condition $a_j^{(n)+}=1$.

Let us introduce the following notation for vector fields 
and quadratic differentials
\begin{equation}
e_k\equiv\psi_{k}^{(-1)},\qquad 
\Omega^k\equiv \psi_{-k}^{(2)}.\label{oiplo}\end{equation}
Note that (\ref{onehalf}) furnishes a basis for the
$1-2s(n)=(2n-1)(h-1)$ holomorphic $n$-differentials
on $\Sigma$
\begin{equation}
{\cal H}^{(n)}=\left\{\psi_k^{(n)}\big |s(n)\le k\le -s(n)\right\}.
\label{qdrtcdfab}\end{equation}
In particular
the quadratic holomorphic differentials are
\begin{equation}
{\cal H}^{(2)}=\left\{\Omega^{k+1-h_0}|k=1,\ldots,(3h-3)\right\},
\qquad h_0\equiv {3\over 2}h.\label{qdrtcdffl}\end{equation}

Let  ${\cal C}$ be a homologically trivial
contour separating $P_+$ and $P_-$.
The dual of $\psi_j^{(n)}$  is defined by
\begin{equation}
{1\over 2\pi i}\oint_{\cal C} \psi_j^{(n)}
\psi^k_{(n)}=\delta_j^k,\label{dualpsi}\end{equation}
which gives
\begin{equation}
\psi^j_{(n)}=\psi_{-j}^{(1-n)}.\label{easy1}\end{equation}

By means of the operator ${\cal S}_{J_H^{-1}}^{(2k+1)}$ 
we define the quantity
\begin{equation}
\chi_F^{(2k+1)}\left(\psi_i^{(-k)},\psi_j^{(-k)}\right)
={1\over 24(2k+1)\pi i}
\oint_{\cal C} \psi_i^{(-k)} {\cal S}_{J_H^{-1}}^{(2k+1)} 
\psi_j^{(-k)},\qquad k=0,1,2,\ldots,\label{ccl}\end{equation}
that for $k=1$ is the Fuchsian $KN$ cocycle
\begin{equation}
\chi_F^{(3)}\left(e_i,e_j\right)
={1\over 24\pi i}
\oint_{\cal C} \left[ {1\over 2}\left( e_ie_j'''-e_i'''e_j\right)+
T^F \left(e_ie_j'-e_i'e_j\right)\right],\quad 
e_j\equiv \psi_j^{(-1)}.\label{cclb}\end{equation}

Notice that 
\begin{equation}
\chi_{f_j}^{(2k+1)}\left(\psi_i^{(-k)},\psi_j^{(-k)}\right)
={1\over 24(2k+1)\pi i}
\oint_{\cal C} \psi_i^{(-k)} {\cal S}_{f_j}^{(2k+1)} 
\psi_j^{(-k)}=0,\quad \forall i,j,
\label{cclg3}\end{equation}
where
\begin{equation}
f_j(z)=\int^z\left[\psi_j^{(-k)}\right]^{-{1\over k}},\quad
{\rm for}\; k=1,2,3,\ldots, \qquad
f_j(z)=\psi_j^{(0)}, \quad {\rm for}\; k=0.
\label{vnsg}\end{equation}

An arbitrary
$KN$ cocycle has the form
\begin{equation}
\widetilde\chi^{(3)}\left(e_i,e_j\right)
=\chi_F^{(3)}\left(e_i,e_j\right)+\sum_{k=1}^{3h-3}a_k
\oint_{\cal C} \Omega^{k+1-h_0}\left[e_i,e_j\right],
\qquad h_0\equiv {3\over 2}h.
\label{cclc}\end{equation}

The cocycle $\widetilde\chi^{(3)}\left(e_i,e_j\right)$ defines
the central extension $\widehat{\cal V}_\Sigma$
 of the $h_0$ graded-algebra ${\cal V}_\Sigma$
of the meromorphic vector fields $\{e_j|j\in{\bf Z}+h/2\}$. In particular
the commutator in $\widehat{\cal V}_\Sigma$ is
\begin{equation}
[e_i,e_j]=\sum_{s=-h_0}^{h_0} C_{ij}^se_{i+j-s}+
t\widetilde \chi^{(3)}\left(e_i,e_j\right),\quad
[e_i,t]=0,\label{knalgeb}\end{equation}
where
\begin{equation}
C_{ij}^s={1\over 2\pi i}\oint_{\cal C}\Omega^{i+j-s}
[e_i,e_j].\label{thecs}\end{equation}
Two important properties of $\widetilde\chi^{(3)}\left(e_i,e_j\right)$
are locality
\begin{equation}
\widetilde\chi^{(3)}\left(e_i,e_j\right)=0,\qquad \hbox{for}\;\,
|i+j|>3h,\label{lcltyh}\end{equation}
and ``time-independence''. Time-independence of $\widetilde\chi^{(3)}$
means that the contribution
to the cocycle is due only to the residue of the integral at the point
$P_+$ ($\tau=0$) or, equivalently at $P_-$ ($\tau=\infty$)
(here we are considering $\tau=e^t$ where $t$ is the time
parameter introduced by Krichever and Novikov which
parametrizes the position of the contour ${\cal C}$ on $\Sigma$).

Let us expand $T^F$ in terms of the $3h-3$ holomorphic 
differentials
\begin{equation}
T^F=\{J_H^{-1},z\}={\cal T}_\Sigma+\sum_{k=1}^{3h-3}
\lambda_k^{(F)}\Omega^{k+1-h_0},\label{ccld}\end{equation}
where ${\cal T}_\Sigma$ denotes the holomorphic projective 
connection on $\Sigma$ obtained from the symmetric differential of the
second-kind with bi-residue 1 and zero $\alpha$-periods
(see \cite{fay} for the explicit expression of ${\cal T}_\Sigma$).

In the case of Schottky uniformization we have
\begin{equation}
T^S=\{J_\Omega^{-1},z\}={\cal T}_\Sigma+\sum_{k=1}^{3h-3}
\lambda_k^{(S)}\Omega^{k+1-h_0},\label{ccld2}\end{equation}
where $J_\Omega:\Omega\to \Sigma$, with $\Omega\subset{\bf\widehat C}$
the region of discontinuity of the Schottky group.
The constants $\lambda_k^{(F)}$
and $\lambda_k^{(S)}$ are the (higher genus)
Fuchsian and Schottkian accessory parameters. 
The Schottkian cocycle is 
\begin{equation}
\chi_S^{(2k+1)}\left(\psi_i^{(-k)},\psi_j^{(-k)}\right)
={1\over 24(2k+1)\pi i}
\oint_{\cal C} \psi_i^{(-k)} {\cal S}_{J_\Omega^{-1}}^{(2k+1)} 
\psi_j^{(-k)},\qquad k=0,1,2,\ldots,\label{ccl8}\end{equation}
that for $k=1$ reduces to the Schottkian $KN$ cocycle
\begin{equation}
\chi_S^{(3)}\left(e_i,e_j\right)
={1\over 24\pi i}
\oint_{\cal C} \left[ {1\over 2}\left( e_ie_j'''-e_i'''e_j\right)+
T^S \left(e_ie_j'-e_i'e_j\right)\right].
\label{cclb3}\end{equation}

The choice of the $KN$ cocycle 
fixes
the normal ordering of operators in higher genus \cite{kn}.
In particular the normal ordering
associated to $\chi_F^{(3)}\left(e_i,e_j\right)$
and $\chi_S^{(3)}\left(e_i,e_j\right)$
depends on the accessory parameters $\lambda_k^{(F)}$
and $\lambda_k^{(S)}$ respectively.
On the other hand these parameters are related to
$S_{cl}^{(h)}$  which denotes the 
Liouville action evaluated on the classical solution \cite{0}. 

To write down $S^{(h)}$ we must consider
the Schottky covering of $\Sigma$. In this approach the relevant group is the
Schottky group ${\cal G}\subset PSL(2,{\bf C})$.
Let $L_1,\ldots,L_h$ be a system of generators for ${\cal G}$
of rank $h>1$ and ${\cal D}$ a fundamental region in the region of
discontinuity $\Omega\subset \widehat {\bf C}$ of $\cal G$ bounded by $2h$
disjoint Jordan curves ${\cal C}_1,{\cal C}'_1,\ldots,{\cal C}_h,
{\cal C}_h'$ such that
${\cal C}_i'=-L_i({\cal C}_i)$. These curves correspond to a cutting of
$\Sigma \cong\Omega/{\cal G}$ along the $\alpha$-cycles.
The Liouville action has the form \cite{0}
$$
S^{(h)}=\int_{\cal D} d^2z
(\partial_z\varphi\partial_{\bar z}{\varphi}+\exp{\varphi})-
{i\over 2}\sum_{i=2}^h\int_{{\cal C}_i}\varphi\left({{\overline L_i''}
 \over{\overline L_i'}}d\bar z-
{L_i''\over L_i'}dz\right)+$$
\begin{equation}
+{i\over 2}\sum_{i=2}^h\int_{{\cal C}_i}{\log}|L_i'|^2{{\overline L_i''}
 \over {\overline L_i'}}d\bar z+
4\pi\sum_{i=2}^h{\log}\left|{(1-\lambda_i)^2
 \over \lambda_i(a_i-b_i)^2}\right|,\label{la1}\end{equation}
where $a_i,b_i\in{\widehat{\bf C}}$ are the attracting and
 repelling fixed points of $L_i$ (it is possible to assume that $a_1=0$, 
$a_2=1$ and $b_1=\infty$) while $\lambda_i$ is defined by the normal form
 \begin{equation}
{L_iz-a_i\over L_iz-b_i}=\lambda_i{z-a_i\over z-b_i},
 \qquad 0<|\lambda_i|<1 .\label{la2}\end{equation}

We now quote the main results in \cite{0}.
The first one concerns the quadratic holomorphic
differential $\Omega=T^F-T^S$ considered as a 1-form 
on Schottky space ${\cal S}$. It turns out that
\begin{equation}
\Omega={1\over 2}\partial 
S_{cl}^{(h)},\label{os0}\end{equation}
where $\partial$ is the holomorphic component of the
exterior differentiation operator on ${\cal S}$.
Furthermore
\begin{equation}
\lambda_k^{(F)}-\lambda_k^{(S)}={1\over 2}{\partial S_{cl}^{(h)}
\over \partial z_k},\qquad k=1,\ldots,3h-3,\label{onp1}\end{equation}
where $\{z_k\}$ are the
coordinates on ${\cal S}$. Another result in \cite{0} is
\begin{equation}
{1\over 2}\overline\partial\partial
S_{cl}^{(h)}=-i\omega_{WP}.\label{os1}\end{equation}
where $\omega_{WP}$ is the Weil-Petersson 2-form on
${\cal S}$.
Furthermore, since $\overline\partial T^S=0$, it follows that
\begin{equation}
\overline\partial T^F=
-i\omega_{WP}.\label{icjdga}\end{equation}

{} From the above results it follows that
the difference between Fuchsian and Schottkian 
higher order cocycles 
depends on the classical Liouville action. In particular for the $KN$ 
cocycle we have 
\begin{equation}
 \chi_F^{(3)}\left(e_i,e_j\right)- 
\chi_S^{(3)}\left(e_i,e_j\right)=
{1\over 48\pi i}
 \sum_{k=1}^{3h-3}{\partial S^{(h)}_{cl}\over 
\partial {z_k}}\oint_{\cal C}\Omega^{k+1-h_0}
\left[e_j,e_i\right].\label{pdop9}\end{equation}
Similar relations hold for the 
Virasoro algebra on punctured Riemann spheres.
Eq.(\ref{pdop9}) clarifies how classical Liouville theory is 
connected with quantum aspects of operators defined on
Riemann surfaces. Let us notice that (\ref{pdop9})
can be generalized to the case of higher order cocycles. Also in this case
the difference between Fuchsian and Schottkian cocycles 
depends on the Liouville action. 

The investigation above solves the problem
posed in \cite{npb} about time-independence and locality 
of the cocycles defined
by covariantization. This follows from the fact that, since the divisor of 
the vector field $\displaystyle {1\over {J_H^{-1}}'}$ is empty,
the integrand in (\ref{ccl}) has no poles outside $P_\pm$.

\subsection{Diffeomorphism Anomaly And The $KN$ Cocycle}

Let us now consider the chirally split form 
of the diffeomorphism anomaly \cite{lazstra}
\begin{equation}
{\cal A}(\mu;e)+\overline{{\cal 
A}(\mu;e)},\label{anomdiff}\end{equation}
where
\begin{equation}
{\cal A}(\mu;e)={1\over 24\pi}
\int_\Sigma\left[{1\over 2} \left(e\partial_z^3\mu-\mu\partial_z^3 e\right)+
T\left(e\partial_z\mu-\mu\partial_z e)\right)\right],
\label{btrlm6}\end{equation}
with  $e$  a vector field. Here $T$ denotes an arbitrary projective 
connection. We now show that ${\cal A}(\mu;e)$ reduces to the $KN$ cocycle.
To do this we first introduce some results on the deformation
of the complex structure of Riemann surfaces \cite{cmp}.
For a short introduction to this subject and related topics
see for example \cite{cn}; for more details see \cite{nit,dhph}. 

To parametrize different
metrics we consider Beltrami differentials
with discontinuities
along a closed curve.
Let $P_+$ be a distinguished point of $\Sigma$ 
and $z_+$ a local coordinate such that
 $z_+(P_+)=0$. Let us denote by  $\Sigma^+$ the disc defined by
$z_+\le 1$, and by $A\subset\Sigma^+$ an annulus whose centre is $P_+$. 
Let $\Sigma^- $ be the surface defined by  
\begin{equation}
\Sigma^+ \cup \Sigma^-  =\Sigma,\qquad
\Sigma^+\cap \Sigma^-  = A.
\label{dowja}\end{equation}
We now perform a change of coordinate
\begin{equation}
z_+\to Z=z_+ + \epsilon e_k(z_+), \qquad z_+\in A,\qquad \epsilon \in 
{\bf C},
\label{h11}\end{equation}
with $e_k\equiv \psi_k^{(-1)}$ a $KN$ vector field.
Identifying the new annulus with the previous collar on $\Sigma^+$ we get
a new surface $\widetilde\Sigma$ whose metric reads
\begin{equation}
 g(\mu_k) =\rho(z,\overline z)
|dz+\mu_k d\overline z|^2,
\label{h12}\end{equation}
where the Beltrami differential is
\begin{equation}
\mu_k(P)=\left\{\begin{array}{ll}
 \epsilon\partial_{\bar z} e_k, & {\rm if} \;\, P\in \Sigma^+;\\
0, &
{\rm otherwise}.\end{array}\right.
\label{h13}\end{equation}

The $KN$ holomorphic differentials $\Omega^j$ form a dual
basis with respect to $\mu_k$. Indeed integrating by parts
we have
\begin{equation}
{1\over \pi}\int_{\Sigma}
\Omega^j\mu_k=\epsilon\delta_k^j.\label{h14}\end{equation}
Since $e_k\sim z_\pm^{\pm k - h_0+1}+\ldots$, it
follows that for $k\ge h_0$ we only change the coordinate $z_+$, whereas
$e_{h_0-1}$ ($h_0\equiv 3h/2$) changes $z_+$ and moves $P_+$.
For $k\le -h_0+1$, $e_k$ is holomorphic on
$\Sigma\backslash\{P_+\}$, so $\widetilde\Sigma$ is isomorphic to
$\Sigma$ because the variation induced in the annulus
can be reabsorbed in a
 holomorphic coordinate transformation
on $\Sigma\backslash \Sigma^+$.
For $|k|\le h_0-2$ the vector field $e_k$ has poles both
in $P_+$ and $P_-$. This change in $\Sigma$
corresponds to an infinitesimal moduli deformation.
Notice that the dimension of the space of these vector fields
is just $3h-3$.

We are now ready to show that the 
anomaly ${\cal A}(\mu;e)$ reduces to the $KN$ cocycle.
First of all notice that by choosing (\ref{h13})
for the Beltrami differential in (\ref{btrlm6}), 
the domain of the surface integral
(\ref{btrlm6}) reduces to $\Sigma^+$. Then we write
${\cal A}(\mu_k;e)$ in the useful form
\begin{equation}
{\cal A}(\mu_k;e)={\epsilon\over 24 \pi}\int_{\Sigma^+}
{e\over v}\partial_z\bigg( v
\partial_z\bigg( v
\partial_z\partial_{\bar z}\bigg({e_k\over v}\bigg)
\bigg)\bigg),\label{equ0}\end{equation}
where $v$ satisfies the equation
\begin{equation}
{1\over 2}\left(v'\over v\right)^2-{v''\over 
v}=T,\label{ohswop}\end{equation}
that for $T=T^F$ has solution
\begin{equation}
v={1\over {J_H^{-1}}'}.
\label{solutionofpre}\end{equation}

We now use the univalence of $J_H^{-1}$. Indeed this guarantees 
that the obstruction for the reduction of (\ref{equ0})
to a contour integral around $\partial \Sigma^+$
(which is homologically equivalent to the ${\cal C}$-contour in 
(\ref{dualpsi})) comes only from possible poles of $e$ in
$\Sigma^+$. As we have seen the univalence of $J_H^{-1}$
implies the holomorphicity of ${\cal S}_{J_H^{-1}}^{(2k+1)}$.
It has been just this property of ${\cal S}_{J_H^{-1}}^{(2k+1)}$ which
has suggested to write the integrand in  ${\cal A}(\mu_k;e)$ in
the form (\ref{equ0}).

Since any diffeomorphism can be expressed in terms of the $KN$ vectors $e_j$, 
it is sufficient to consider ${\cal A}(\mu_k;e_j)$ instead of 
${\cal A}(\mu_k;e)$.
By the remarks above it follows that\footnote{Note that for $T\ne  T^F$ 
we have a similar relation.}
\begin{equation}
{\cal A}(\mu_k;e_j)=
{\epsilon\over 2} \chi^{(3)}_F(e_j,e_k), \qquad j\ge h_0-1.
\label{hgt}\end{equation}
For $j\le h_0-2$, the vector field $e_j$ has poles at $z=P_+$ and 
${\cal A}(\mu_k;e_j)$ can be expressed as a linear combination of 
$KN$ cocycles.

Note that the Wess-Zumino condition for 
${\cal A}(\mu_k;e_j)$ corresponds to the cocycle identity 
for $\chi^{(3)}_F(e_j,e_k)$. On the other hand 
writing
$\chi^{(3)}_F(e_j,e_k)$ in terms of theta functions 
(the explicit form of $e_k\equiv\psi_k^{(-1)}$ is given in (\ref{psij}))
 one should get some constraints on the period matrix
{}from  the cocycle identity that
presumably are connected
with the Hirota bilinear relation. Thus the Wess-Zumino
condition for ${\cal A}(\mu_k;e_j)$ 
seems to be related to the Schottky problem.
We do not perform such analysis here, however we stress that 
the cocycle condition for $\chi^{(3)}_F(e_j,e_k)$ involves,
besides the period matrix, the Fuchsian accessory parameters.

Eq.(\ref{hgt}) suggests to define the higher order anomalies 
\begin{equation}
{\cal A}\left(\mu_j^{(2k+1)},\psi_i^{(-k)}\right)
={1\over 24(2k+1)\pi }
\int_{\Sigma} \psi_i^{(-k)} {\cal R}_{J_H^{-1}}^{(2k+1)} 
\psi_j^{(-k)},\qquad k=0,1,2,\ldots,\label{ccaal}\end{equation}
with
\begin{equation}
{\cal R}_{f}^{(2k+1)} =(2k+1)
(f')^k \partial_z (f')^{-1}\partial_z (f')^{-1}\ldots
\partial_z (f')^{-1}\partial_z\partial_{\bar z} (f')^k,
\label{cvprtar}\end{equation}
where the number of derivatives is $2k+1$. 
Notice that the generalized Beltrami differentials are
\begin{equation}
\mu^{(2k+1)}_k(P)=\left\{\begin{array}{ll}
 \epsilon\partial_{\bar z} \psi_j^{(-k)}, & {\rm if} \;\, P\in \Sigma^+;\\
0, &
{\rm otherwise}.\end{array}\right.
\label{h13aa}\end{equation}
Here the deformation of the complex structure
of vector bundles on Riemann surfaces is provided by the space 
of differentials
\begin{equation}
\widetilde{\cal H}^{(k+1)}=\left\{\psi_j^{(-k)}\big|1-s(-k)\le j
\le s(-k)-1\right\},
\label{qdrtcdfabc}\end{equation}
which is the dual space to ${\cal H}^{(k+1)}$ defined in 
(\ref{qdrtcdfab}).

By construction higher order anomalies are related to higher order cocycles
in a way similar to eq.(\ref{hgt}). In this case we 
must consider ${W}$-algebras and the moduli space of vector bundles on 
Riemann surfaces. We notice that the explicit expression of the 
$KN$-differentials in terms of theta functions given in (\ref{psij}) provides 
a useful tool to investigate this subject.

We now show that in the case one uses the generalized
harmonic Beltrami differentials
\begin{equation}
\mu_{harm,\,j}^{(2k+1)}=\overline\psi^{(k+1)}_je^{-k\varphi},\qquad
j\in [s(k+1),-s(k+1)],\; s(k)=h/2-(k)(h-1),
\label{iudcx}\end{equation}
the anomalies (including the standard chirally split
anomaly) vanish
\begin{equation}
{\cal A}\left(\mu_{harm,\,j}^{(2k+1)},\psi_i^{(-k)}\right)
={1\over 24(2k+1)\pi}
\int_{\Sigma} \psi_i^{(-k)} {\cal S}_{J_H^{-1}}^{(2k+1)} 
\mu_{harm,\,j}^{(2k+1)}=0,\qquad k=0,1,2,\ldots.\label{cc1aal}\end{equation}
This follows simply because
\begin{equation}
{\cal S}^{(2k+1)}_{J_H^{-1}}\mu_{harm,\, j}^{(2k+1)}=0.
\label{gcex}\end{equation}

\mysection{Virasoro Algebra On $\Sigma$}

Here we consider a sort of higher genus generalization
of the Killing vectors. This generalization follows from an
investigation of the kernel of the $KN$ cocycle. 
This analysis will suggest 
two possible realizations of the 
Virasoro algebra on $\Sigma$ without central extension
based on the Poincar\'e metric and $J^{-1}_H$.
Finally a higher genus realization of the Virasoro cocycle 
$\chi_{kj}=\delta_{k,-j}(j^3-j)/12$ is given.

\subsection{Higher Genus Analogous Of Killing Vectors}

In genus zero the $KN$ cocycle reduces to $\chi_{kj}$ which vanishes
for $j=-1,0,1$, $\forall k$. This reflects the $SL(2,{\bf C})$
symmetry of the Riemann sphere due to the three Killing vectors.
For $h\ge 2$ do not exist chiral holomorphic
$-k$-differentials, with $k=1,2\ldots$. The reason is that
in this case 
\begin{equation}
{\rm deg}\,\psi_j^{(-k)}=2k(1-h)<0,
\label{lidlkj}\end{equation}
so that $\psi_j^{(-k)}$ has at least $2k(h-1)$ poles. Nevertheless by 
eq.(\ref{ocihwe1}) the cocycles have the following property
\begin{equation}
\chi_F^{(2k+1)}\left(\psi_i^{(-k)},
\left(2\varphi_{\bar z}\right)^le^{-k\varphi}\right)=0,\qquad 
\forall i, \qquad k=0,1,2,\ldots,\quad l=0,\ldots,2k.
\label{foralli7}\end{equation}
In particular 
\begin{equation}
\chi_F^{(3)}\left(e_i,e^{-\varphi}\right)=
\chi_F^{(3)}\left(e_i,2\varphi_{\bar z}e^{-\varphi}\right)=
\chi_F^{(3)}\left(e_i,(2\varphi_{\bar z})^2e^{-\varphi}\right)=
0,\qquad \forall i.
\label{foralli0}\end{equation}
Thus, in spite of the fact that for $h\ge 2$ Killing vectors do not 
exist, the non-chiral vectors $(2\varphi_{\bar z})^le^{-\varphi}$,
$l=0,1,2$, can be seen as their higher genus 
generalization. Let us make some remarks on this point.
The Killing vectors are the solutions of the equation
\begin{equation}
\partial_{\bar z} v=0.\label{killingsp}\end{equation}
In the case of the Riemann sphere we 
can choose the standard atlas $(U_\pm,z_\pm)$ with $z_-=z_+^{-1}$
in the intersection 
$U_+\cap U_-$. For the component of a vector field 
$v\equiv \{v^+(z_+),v^-(z_-)\}$, we 
have $v^+(z_+)(dz_+)^{-1}= v^-(z_-)(dz_-)^{-1}$, that is 
$v^-(z_-)=-z^2_-v^+(z_-^{-1})$. Therefore
if 
\begin{equation}
v_l^+(z_+)=z_+^l,\label{hclk}\end{equation}
then $v_l^-(z_-)=-z^{-l+2}_-$ and
the solutions of eq.(\ref{killingsp}) are $v_0,v_1,v_2$.
To understand what happen in higher genus,
we first note that besides  eq.(\ref{killingsp})
these vector fields are solutions of
the covariant equation
\begin{equation}
{\cal S}_f^{(3)}\cdot v=0,\qquad f(z)\equiv \int^z v_0^{-1}.
\label{podj}\end{equation}
Indeed in $U_+\cap U_-$ eq.(\ref{podj}) reads
\begin{equation}
\partial^3_{z_+}v^+(z_+)=
z_-^{2}
\partial_{z_-}\bigg(z_-^2\partial_{z_-}\bigg(z_-^2\partial_{z_-}
\bigg(z_-^{-2}
v^-(z_-)\bigg)\bigg)\bigg)=0,\label{hfgto}\end{equation}
whose solutions coincide with the solutions 
of eq.(\ref{killingsp}). This relationship between
the  zero modes of $\partial_{\bar z}$ and  
${\cal S}_f^{(3)}$
extends to the case of $-k$-differentials, $k=0,1/2,1,\ldots$.
In particular on the Riemann sphere the $2k+1$ 
chiral solutions of the
equation ${\cal S}_f^{(2k+1)}\cdot \phi^{(-k)}=0$, 
where $\phi^{(-k)}$ are $-k$-differentials, coincide with the
zero modes of the $\partial_{\bar z}$ operator
\begin{equation}
{\cal S}_f^{(2k+1)}\cdot \phi^{(-k)}=0
\quad \longrightarrow\quad  \partial_{\bar z}\phi^{(-k)}=0, 
\qquad k=0,{1\over 2},1,\ldots,
\label{correspn}\end{equation}
whose solutions are $\phi_l^{(-k)}\equiv\{\phi_l^{(-k)+},
\phi_l^{(-k)-}\}$, $l=0,1,2$, where 
\begin{equation}
\phi_l^{(-k)+}(z_+)=z_+^l,\quad\qquad 
\phi_l^{(-k)-}(z_-)=(-1)^k z_-^{2k-l}.\label{uslp}\end{equation}
The higher genus generalization of (\ref{correspn}) 
reads
\begin{equation}
{\cal S}_{J^{-1}_H}^{(2k+1)}\cdot
\phi^{(-k)}=0 \quad \longrightarrow\quad
\partial_{\bar z} \phi^{(-k)}=0,
\qquad k=0,{1\over 2},1,\ldots,
\label{kgfhfuid}\end{equation}
whose solutions are\footnote{However note that $\partial_{\bar z} 
\phi^{(-k)}_l=0,\forall l$.} 
\begin{equation}
\phi_l^{(-k)}={\left(J^{-1}_H\right)^l\over \left({J^{-1}_H}'\right)^{k}},
\qquad l=0,1,\ldots,2k.\label{fdgsts}\end{equation}
Thus a possible choice for the higher genus analogous
of the Killing vectors are the polymorphic vector fields
\begin{equation}
\phi_0^{(-1)}={1\over {J^{-1}_H}'},\qquad
\; \phi_1^{(-1)}={J^{-1}_H\over {J^{-1}_H}'},
\qquad \; \phi_2^{(-1)}
={\left(J^{-1}_H\right)^2\over {J^{-1}_H}'}.
\label{jxd}\end{equation}
Similarly to the case of the Killing vectors, $\phi_0^{(-1)}$, 
$\phi_1^{(-1)}$ and $\phi_2^{(-1)}$
are zero modes for $\chi^{(3)}_F$. More generally
\begin{equation}
\chi_F^{(2k+1)}\left(\psi_i^{(-k)},
\phi_l^{(-k)}\right)=0,\qquad 
\forall i, \qquad k=0,1,2,\ldots,\quad l=0,\ldots,2k.
\label{foralli9}\end{equation}

However if singlevaluedness is required we must relax the chirality
condition and instead of $\phi_l^{(-k)}$ we must
consider the non chiral differentials
$(2\varphi_{\bar z})^le^{-k\varphi}$,
$l=0,1,\ldots,2k$.

\subsection{Realization Of The Virasoro Algebra On $\Sigma$}

The previous discussion suggests a higher genus realization
of the Virasoro algebra.
We first consider two realizations of this algebra without
central extension.
In the first case we have
\begin{equation}
\left[L_j,L_k\right]
=(k-j)L_{j+k}, \qquad L_k=(2\varphi_{\bar z})^{k+1}e^{-\varphi}\partial_z.
\label{virsi1}\end{equation}

Similarly we can realize the centreless Virasoro algebra on $\Sigma$
considering as generators the polymorphic chiral vector fields
\begin{equation}
L_k^{ch}={\left(J^{-1}_H\right)^{k+1}\over{J^{-1}_H}'}\partial_z,
\quad \qquad \overline 
L_k^{ch}={\left(\overline J^{-1}_H\right)^{k+1}\over{\overline J^{-1}_H}'}
\partial_{\bar z},
\label{odilpk}\end{equation}
so that
\begin{equation}
\left[L_j^{ch},L_k^{ch}\right]
=(k-j)L_{j+k}^{ch}, \qquad \left[\overline L_j^{ch},
\overline L_k^{ch}\right]
=(k-j)\overline L_{j+k}^{ch},\qquad 
\left[L_j^{ch},
\overline L_k^{ch}\right]=0.
\label{virsi2}\end{equation}

Observe that the holomorphic operators
${\cal S}_{J^{-1}_H}^{(2k+1)}$ can be expressed in terms of the
above generators
\begin{equation}
{\cal S}_{J^{-1}_H}^{(2k+1)}=(2k+1) \left({J_H^{-1}}'\right)^{k+1}
{L_{-1}^{ch}}^{2k+1}\left({J_H^{-1}}'\right)^k=
(2k+1)e^{(k+1)\varphi} L_{-1}^{2k+1}e^{k\varphi}.
\label{bobafd}\end{equation}

The structure of the generators $L_k^{ch}$ suggests the
generalization
\begin{equation}
{\cal L}_k=v_k\partial_z,\qquad v_k(z)={f^{k+1}(z)\over f'(z)},
\label{callp}\end{equation}
with $f(z)$ an arbitrary meromorphic function. In this case we can
define the cocycle
\begin{equation}
\chi(v_k,v_j)={1\over 24\pi i}
\oint_{{\cal C}_0} v_k {\cal S}^{(3)}_f v_j=
{j^3-j\over 12}\delta_{k,-j},
\label{newcoyc}\end{equation}
where ${\cal C}_0$ encircles a simple zero of $f$. Thus we have
\begin{equation}
\left[{\cal L}_j,{\cal L}_k\right]
=(k-j){\cal L}_{j+k} + {j^3-j\over 12}\delta_{k,-j}.
\label{virsiab1}\end{equation}

To define a cocycle depending only on the homological class of the 
contour we consider
\begin{equation}
f=e^{\int^z \omega},\label{djakj}\end{equation}
with $\omega$ a 1-differential. A possible choice is
to consider the third-kind differential with poles at
$P_\pm$ and with periods over all cycles imaginary
\begin{equation}
\omega=\partial_z\log {E(z,P_+)\over E(z,P_-)}-
2\pi i \sum_{j,k=1}^h\left({\rm Im}\,
\int_{P_-}^{P_+}\omega_j\right){\Omega^{(2)}_{jk}}^{-1}
\omega_k(z),\label{dsjgnv}\end{equation}
where $\Omega^{(2)}$  denotes the imaginary part of the Riemann period 
matrix. In this case one can substitute the contour ${\cal C}_0$ in
(\ref{newcoyc}) with the contour ${\cal C}$ in (\ref{dualpsi}).
However we notice that the integrand of 
(\ref{newcoyc}) with $f$ given in (\ref{djakj}) is not
singlevalued for $j\neq -k$. An interesting
 possibility to investigate this aspect is to set
$f=f_{n,m}$ where the $f_{n,m}$'s are defined in the next section.

\mysection{Liouville Field And Higher Genus Fourier Analysis}

In the standard approach to 2D gravity the Liouville field is 
considered as a free field. However there is a substantial 
hindrance in the CFT approach to Liouville gravity.
Namely, since  the metric $g=e^\sigma \hat g$ must be well-defined,
$e^\sigma$ must be an element of ${\cal C}^\infty_+$.
If $\sigma$ were considered as a free scalar
field then the metric would take non positive values as well. 
One of the aims of this section is to investigate uniformization 
theory and then provide the mathematical tools to face the problem of 
metric positivity in considering Liouville gravity on higher genus 
Riemann surfaces.

A long-standing problem in uniformization theory is to express the 
uniformizing Fuchsian group in terms of the Riemann
period matrix $\Omega$. A 
possibility is to write $J_H^{-1}$ in terms of $\Omega$. In this case,
after going around non trivial cycles, the coefficients in the M\"obius
transformation of $J_H^{-1}$ are given in terms of $\Omega$. 
Unfortunately to write explicitly $J^{-1}_H$ seems to be an 
outstanding problem. 

These aspects are related to the problem of finding the 
eigenfunctions of the Laplacian on $\Sigma$. Actually, there is a strict 
relationship between the eigenvalues of the Laplacian and geodesic 
lenghts. These lenghts are related to the trace of hyperbolic Fuchsian 
elements. A way to investigate this argument is by Selberg trace 
formula. However one should try to investigate the problem of 
uniformization in a more direct (analytic) way. Here we introduce a new set
 of functions on $\Sigma$ whose structure seems related to 
the uniformizing Fuchsian group.

\subsection{Positivity And Fourier Analysis On Riemann Surfaces}

A possible way to construct functions in ${\cal C}^\infty_+$
is to consider 
the ratio of two suitable
$(p,q)$-differentials.
For $p=q=1$, besides the Poincar\'e
metric, we can use 
any positive quadratic form
like $\sum_{j,k=1}^h
\omega_jA_{jk}\overline \omega_k$.
An alternative is to attempt to define the higher genus analogous of the 
Fourier modes. In the following we adopt this approach.

Let
\begin{equation}
G={e^{g-\overline  g}+ e^{\overline g-g}\over 2}
={\rm cos}\, \left(2\,{\rm Im}\, g\right),\label{split}\end{equation}
be a function on  a compact Riemann surface $\Sigma$.
We will see that in order that $G$ be a non trivial regular function
in ${\cal T}^{0,0}$, it is necessary that 
 after winding around the homology cycles of 
$\Sigma$, the function $g$
 transforms with an additive term whose imaginary
part be a non-vanishing element in 
$\pi {\bf Z}$. Such a multivaluedness 
is crucial for the construction of well-defined 
regular functions on $\Sigma$.
We will see that there are infinitely many functions,
labelled by $2h$ integers $(n,m)\in{\bf Z}^{2h}$,
with the properties of $g$ whose existence is strictly related to the
positive definiteness of the imaginary part of the Riemann period matrix.

\subsection{Real Multivaluedness And ${\rm Im}\,\Omega>0$}

We begin by considering the holomorphic differential
 \begin{equation}
\omega(z)=\sum_{k=1}^hA_k\omega_k(z),
\qquad {A}={a}+i{b},\quad ({a,b})
\in {\bf R}^{2h},\label{m21} \end{equation} 
where $\omega_1,\ldots,\omega_h$, are the
holomorphic differentials with the standard normalization (\ref{stndnorm}).

After winding around the cycle
\begin{equation}
 c_{q,p}={p}\cdot{\alpha}+{q}\cdot {\beta}, 
\qquad (q,p)\in {\bf Z}^{2h},\label{babab}\end{equation}
the function $f(z)=e^{\int_{P_0}^z \omega}$
transforms into 
\begin{equation}
f(z+c_{q,p})= 
\exp\left({\sum_{k=1}^h(p_k+\sum_{l=1}^h q_l\Omega_{kl})A_k}\right)
f(z).\label{baba2}\end{equation}
We constrain the multivaluedness factor in 
(\ref{baba2}) to be real for arbitrary $({q,p})\in {\bf Z}^{2h}$;
that is we require that the imaginary part of 
the exponent in (\ref{baba2}) be an integer multiple of $\pi$
\begin{equation}
\sum_{k=1}^hb_k\oint_{\alpha_j}\omega_k=\pi n_j,\qquad j=1,\ldots,h,
\qquad n_j\in{\bf Z},\label{m22}\end{equation}
\begin{equation}
\sum_{k=1}^h  a_k\Omega_{kj}^{(2)} +\sum_{k=1}^h b_k
\Omega_{kj}^{(1)}= \pi  m_j, 
\qquad j=1,\ldots,h, \qquad m_j\in{\bf Z}, \label{m23}\end{equation}
where 
\begin{equation}
\Omega_{kj}^{(1)}\equiv{\rm Re}\,
\Omega_{kj}, \qquad \Omega_{kj}^{(2)}\equiv {\rm Im}\, \Omega_{kj}.
\label{omegaper}\end{equation}
Thus after winding around $\alpha_j$  we have
\begin{equation}
f(z+\alpha_j)=\exp (a_j+i\pi n_j)f(z),
\label{alphaj}\end{equation}
whereas around $\beta_j$ 
\begin{equation}
f(z+\beta_j)=\exp\left [\sum_{k=1}^h \left
 (a_k\Omega^{(1)}_{kj}-\pi n_k\Omega_{kj}^{(2)}
\right)+i\pi m_j\right]
f(z).\label{betaj}\end{equation}
Eqs.(\ref{m22},\ref{m23}) show an interesting connection between
a fundamental property of Riemann surfaces and the existence
of regular functions with real
multivaluedness. Namely,
 for each fixed set  of integers 
$({n,m})\ne ({0,0})$, positivity of
$\Omega^{(2)}$ guarantees the existence
of a non trivial solution
of eqs.(\ref{m22},\ref{m23}). We have
\begin{equation}
a_k={{\rm det}\, \Omega^{(2;k)}\over {\rm det}
\,\Omega^{(2)}}, \quad\qquad b_k=\pi n_k,\qquad\qquad n_k\in{\bf Z},
\label{m24}\end{equation}
where $\Omega^{(2;k)}$ is obtained by the matrix $\Omega^{(2)}$ after the 
substitutions
\begin{equation}
\Omega^{(2)}_{kj}\to \pi \left(m_j-\sum_{l=1}^h n_l
\Omega_{lj}^{(1)}\right),
\qquad\qquad j=1,\ldots,h.\label{m25}\end{equation}
For practical reasons we change the notation of
 $f$ and $\omega$
\begin{equation}
f_{n, m}(z)=\exp \int_{P_0}^z
\omega_{n, m},\quad
\omega_{n, m}(z)=
\sum_{k=1}^h\left({\det \Omega^{(2;k)}\over
\det \Omega^{(2)} }+i\pi n_k\right)\omega_k(z).
\label{facca}
\end{equation}

We now illustrate some interesting
properties of the functions $f_{n,m}$.

\subsection{Eigenfunctions}

Let $({n,m})\in {\bf Z}^{2h}$ be fixed
and consider the scalar Laplacian $\Delta_{g,0}=
-g^{z\bar z}\partial_{z}\partial_{\bar z}$
with respect to the degenerate metric
\begin{equation}
ds^2=2g_{z\bar z}|dz|^2,\qquad
g_{z\bar z}={|\omega_{n,m}|^2\over 2A},
\label{me1}\end{equation}
where $A$ normalizes the area of $\Sigma$ to 1 (see eq.(\ref{area111})).
The functions
\begin{equation}
\psi_k(z,\bar z)=
{1\over \sqrt 2} \left[ \left ( {f_{n, m}(z)\over
\overline  f_{n, m}(z)}\right)^{k}+
\left ( {\overline {f}_{n, m}(z)\over \ f_{n, m}(z)}
\right)^{k}\right],
\qquad k=0,1,2,\ldots,
\label{eigs}\end{equation}
with $(n,m)$ fixed, are eigenfunctions
of $\Delta_{g,0}$ with eigenvalues
\begin{equation}
\Delta_{g,0} \psi_k(z,\bar z)
 = \lambda_k   \psi_k(z,\bar z),\qquad
 \lambda_k = 2A k^2,\qquad k=0,1,2,\ldots .
\label{eqautov}\end{equation}

Note that
\begin{equation}
2A\sum_{k=1}^\infty {1\over \lambda_k}=\zeta (1)={\pi^2\over 6},
\quad\qquad\quad 4A^2\sum_{k=1}^\infty
 {1\over \lambda_k^2}=\zeta (2)={\pi^4\over 90}.
\label{zzz}\end{equation}
The orthonormality of the eigenfunctions
\begin{equation}
\int_\Sigma  \sqrt g \psi_j\psi_k=
\delta_{jk},\label{ort}\end{equation}
follows from the fact that
$|\omega_{n,m}(z)|^2\exp k\left(\int^z\omega_{n,m}
-{\overline {\int^z\omega_{n,m}}}\right)$
is a total derivative.

Let us notice that 
for $k\notin {\bf Z}$ the functions
 $\left(\overline {f}_{n, m}/ f_{n, m}\right)^k$ are in general (see 
below)
not well-defined. 
This shows that arbitrary powers of well-defined
scalar functions can be multivalued around the homology cycles.
In this sense 
the possible values of $k$ in (\ref{eigs}) are fixed by ``boundary 
conditions''.  This aspect should be
taken into account in considering operators
such as $e^{\alpha\phi}$ in Liouville and conformal field theories.

We stress that in general there are other eigenfunctions besides
$\psi_k$, $k\in {\bf N}$. For example
when all the $2m_j$'s and $2n_j$'s are integer multiple
of an integer $N$ then the eigenfunctions include $\psi_{k/N}$,
$k\in {\bf N}$ whose eigenvalue is
$2Ak^2/N^2$.
More generally one should investigate whether the period matrix has some
non trivial  number theoretic structure. 
For example the problem of finding the possible
solutions of the equation 
\begin{equation}
\omega_{n',m'}(z)=c\cdot\omega_{n,m}(z),  
\label{newwsad}\end{equation}
with both $(n,m)$ and $(n',m')$ in ${\bf Z}^{2h}$ and $c$ a 
(in general complex) constant, is strictly related to the numerical
properties of $\Omega$.

On the other hand if eq.(\ref{newwsad}) has non trivial 
solutions (by trivial we mean $c\in {\bf Q}$) then 
\begin{equation}
{f_{n', m'}(z)\over
\overline  f_{n', m'}(z)}
=e^{c\int^z\omega_{n,m}-\overline c\overline{\int^z\omega_{n,m}}}
\neq \left({f_{n, m}(z)\over
\overline  f_{n, m}(z)}\right)^k
\qquad
c\in {\bf C}\backslash {\bf Q},\quad k\in {\bf Q}.
\label{a1a2e3i4g5s}\end{equation}
Therefore
\begin{equation}
\Delta_{g,0}\phi_c(z,\bar z)=2A|c|^2\phi_c(z,\bar z),
\qquad \Delta_{g,0}=-2A|\omega_{n,m}|^{-2}\partial_z\partial_{\bar z},
\label{a1a2e3i4g5s6787}\end{equation}
where
\begin{equation}
\phi_c(z,\bar z)={1\over \sqrt 2} \left({f_{n', m'}(z)\over
\overline  f_{n', m'}(z)}+
{\overline {f}_{n', m'}(z)\over \ f_{n', m'}(z)}
\right).
\label{aaeigs}\end{equation}
The relevance of these functions resides in the
non trivial  number theoretic (chaotic) structure
of the eigenvalues 
\begin{equation}
\lambda_c=2A|c|^2.
\label{iudqdpolk}\end{equation}
To understand this it is sufficient to write 
eq.(\ref{newwsad}) in the 
more transparent form
\begin{equation}
m_j'-\sum_{k=1}^h\Omega_{jk}n_k'=\overline c
\left(m_j-\sum_{k=1}^h\Omega_{jk}n_k\right),\qquad j=1,\ldots,h.
\label{iudqwwww}\end{equation}
To each $(n,m)$ and $(n',m')$ satisfying (\ref{iudqwwww}) corresponds 
a possible value of $c$.

Since (\ref{newwsad}) (or equivalently (\ref{iudqwwww}))
are equations on $\Omega$
one should think that they are related to the 
Hirota bilinear relations or equivalently to the Fay trisecant
identity\footnote{Recall that the Fay trisecant
identity is the higher genus generalization of the Cauchy 
(= bosonization) formula 
$${\prod_{i<j}(z_i-z_j)(w_i-w_j)\over \prod_{ij}(z_i-w_j)}=
(-1)^{n(n-1)/2}\det\left({1\over z_i-w_j}\right).$$}. 
As well-known these must be satisfied by 
$\Omega$ (Schottky's problem). These remarks indicate that the 
aspects of number theory underlying the structure of Fuchsian groups 
should be related to integrable systems such as the KP hierarchy.

The construction of the functions 
$h_{n,m}=f_{n,m}/{\overline f}_{n,m}$ indicates that the solutions of the
equation
\begin{equation}
\partial_z \psi(z,\bar z)=\mu(z)\psi(z,\bar z),
\label{cwc}\end{equation}
with $\mu$ a holomorphic differential and $\psi$ a smooth function are
$\mu=\omega_{n,m}$ and $\psi=h_{n,m}$. 
It seems that equations such as (\ref{newwsad}) and (\ref{cwc})
provide analytic tools 
to investigate aspects 
of number theory, structure of moduli space, chaotic spectrum etc.

\subsection{Multivaluedness, Area And Eigenvalues}

To evaluate $A$ we can use
the Riemann bilinear relations
\begin{equation}
2A= {i\over 2} \int_\Sigma \omega_{n,m}\wedge \overline\omega_{n,m}
=-{\rm Im}
\sum_{j=1}^h\oint_{\alpha_j} \omega_{n,m}\oint_{\beta_j}\overline
\omega_{n,m}=
\sum_{l,k=1}^h (a_ka_l+b_kb_l)\Omega_{lk}^{(2)}, \label{bob1}\end{equation}
and by  eqs.(\ref{m22},\ref{m23})
\begin{equation}
A={\pi\over 2}\sum_{l=1}^h\left\{a_l\left(m_l-\sum_{k=1}^h
\Omega_{lk}^{(1)}n_k\right)+\pi
\sum_{k=1}^hn_l\Omega_{lk}^{(2)}n_k\right\}.
\label{area111}\end{equation}
The multivaluedness of $f_{n, m}$ is related to $A$ (the
area of the metric $|\omega_{n,m}(z)|^2$). In particular,
after winding around the cycle
$c_{n,-m}=-{m} \cdot {\alpha}+ {n} \cdot {\beta}$, we have
\begin{equation}
{\cal P}_{n,-m}f_{n, m}(z)=e^{-{2 A\over \pi}} 
f_{n, m}(z),
\label{trmn}\end{equation}
where ${\cal P}_{q,p}$ is the winding operator
\begin{equation}
{\cal P}_{q,p}g(z)=g(z+c_{q,p}).\label{wndngprtr}\end{equation}

Comparing (\ref{trmn}) with (\ref{eqautov}) we get the following 
relationship connecting multivaluedness, area and eigenvalues
\begin{equation}
\lambda_k={1\over \pi}\log{ f_{n, m}(z)\over
f_{n, m}(z+k^2c_{n,-m})}.\label{eigmulti}\end{equation}
Thus we can express the action of the Laplacian in terms of the 
winding operator. This relationship between eigenvalues 
and multivaluedness is reminiscent of a similar relation
arising between geodesic lenghts (Fuchsian dilatation) and eigenvalues
of the Poincar\'e Laplacian (Selberg trace formula).

\subsection{Genus One}

One of the properties of the $f_{n,m}$'s is that
in the case of the torus the functions
\begin{equation}
\phi_{n,m}(z,\bar z)=  {1\over \sqrt 2}\left({f_{n,m}(z)\over \overline  
{f}_{n,m}(z)}+{\overline f_{n,m}(z)\over {f}_{n,m}(z)}\right), 
\qquad\qquad\;({n,m})\in {\bf Z}^2,\label{torus2}\end{equation}
coincide with the  well-known eigenfunctions for
the Laplacian $-2\partial_z\partial_{\bar z}$.
To prove this we choose the
 coordinate $z=x+\tau y$ with
 $\tau=\tau^{(1)} +i\tau^{(2)}$ the torus period matrix.
Eq.(\ref{m24}) gives
\begin{equation}
a= {\pi({m-n}\tau^{(1)})\over \tau^{(2)}},
\qquad b=\pi {n},\label{torus}\end{equation}
thus, choosing $P_0=0$, we get
\begin{equation}
\phi_{n,m}(z,\bar z) =\sqrt 2 {\rm cos}\, 2\pi (nx +my),
\qquad (n,m)\in {\bf Z}^2,
\label{torus3}\end{equation}
and
\begin{equation}
\lambda_{n,m}=
{2\pi^2}({m}-\tau {n})({m}-\overline \tau {n})/{\tau^{(2)}}^2,
\qquad (n,m)\in {\bf Z}^2.
\label{toruss}\end{equation}

\subsection{Remarks}

Let us make further remarks about $h_{n,m}={f_{n, m}/\overline  f_{n, m}}$ 
in (\ref{eigs}). First of all note
that in considering $h_{n,m}^k$ ($=h_{k{n},k{m}}$)
as eigenfunctions of the scalar 
Laplacian, the indices $(n,m)$ are fixed whereas in the case of the 
torus the eigenfunctions are $h_{n,m}$ with 
$(n,m)$ running in ${\bf Z}^2$. Thus if we insist on using
$h_{n,m}^k$ with fixed $({n,m})$ also on the torus,
then we will lose infinitely many eigenfunctions of the Laplacian
$-2\partial_z\partial_{\bar z}$. Therefore, for analogy with the torus 
case,
a complete set of eigenfunctions in higher genus should be labelled
by $({n,m})\in {\bf Z}^{2h}$.
However the structure of eq.(\ref{iudqwwww}) 
suggests that there is a constraint
on the values of $(n,m)$ which should reduce the number of indeces to $h$.

Unfortunately it is very difficult 
to recognize a complete set of eigenfunctions in higher 
genus.
This question is related to the problem of finding the 
explicit dependence of the Poincar\'e metric $e^\varphi$ on the 
moduli\footnote{As we have seen this would be equivalent to 
finding the 
explicit dependence of $J_H^{-1}$ on the moduli of $\Sigma$
and then to solving long-standing 
problems in the theory of uniformization, Fuchsian groups etc..}.
The reason is that if in the torus case
the complete set of eigenfunctions should reduce 
to $\{\phi_{n,m}\}$ then for analogy the Laplacian on
higher genus surfaces must be definite with respect to the 
constant curvature metric, that is the Poincar\'e 
metric. Really, each metric
in the form $g_{z\bar z}^{(p)}=\sum_{j,k=1}^h
\omega_jA_{jk}^{(p)}\overline \omega_k$, with $A^{(p)}$ a
positive definite matrix, 
reduces to the constant curvature metric on the torus.
However ${\rm det}'\, \Delta_{g^{(p)},0}$
and ${\rm det}'\, \Delta_{g^{(q)},0}$ are related by the Liouville action for
the Liouville field $\sigma=\log g_{z\bar z}^{(p)}/g_{z\bar z}^{(q)}$
in the background metric $g^{(q)}$.

Notice that starting from the
requirement of real multivaluedness, 
that is the ``Dirac condition'' (\ref{m22},\ref{m23}), 
we end in a natural way with a set of functions 
with important properties. In particular, since this condition
is the basic feature underlying the construction of eigenfunctions in the
case of the torus, it seems that the Dirac condition is a guidance
to formulate Fourier analysis on higher genus Riemann surfaces as well.
Thus the properties of the $f_{n,m}$'s suggest that
they are a sort of
``building-blocks'' to
construct a complete set of eigenfunctions for the scalar Laplacian
of a well-defined metric. In particular the set of real functions 
\begin{equation}
{\cal F}=\left\{{\rm cos}\,\left( 2{\rm Im} g_{n,m}(z)\right),\,
{\rm sin}\,\left( 2{\rm Im} g_{n,m}(z)\right)
\bigg| ({n,m})\in {\bf Z}^{2h}\right\}, \quad g_{n,m}(z)\equiv 
\int_{P_0}^z\omega_{n, m},
\label{pf}\end{equation}
 resemble higher genus Fourier modes.
Furthermore, by the analogy with
the torus case, one should investigate whether
\begin{equation}
\lambda_{n,m}= 2\sum_{l,k=1}^h (a_ka_l+b_kb_l), \qquad 
(n,m)\in {\bf Z}^{2h},
\label{eigenvls}\end{equation}
are eigenvalues of the Laplacian with respect to some metrics.
Note that
the term $a_ka_l+b_kb_l$ in (\ref{eigenvls}) appears in the expression
for the area of the metric $|\omega_{n,m}(z)|^2$ (see (\ref{bob1})).
On the other hand $\Omega^{(2)}/\det \Omega^{(2)}$ reduces
to 1 in genus 1, therefore still by analogy with the torus case
one should consider
possible candidates for eigenvalues also the following quantities
\begin{equation}
\mu_{n,m}= {2\over \det \Omega^{(2)}}
\sum_{l,k=1}^h (a_ka_l+b_kb_l)\Omega^{(2)}_{lk}=
{4 A_{n,m}\over \det \Omega^{(2)}}, \qquad 
(n,m)\in {\bf Z}^{2h},
\label{e1i2g3e4n5v6l7s89}\end{equation}
with $A_{n,m}\equiv A$ where the dependence of the area $A$ 
on $(n,m)$ is given in (\ref{area111}).
Other quantities that should be evaluated are
\begin{equation}
{\cal Z}_h(\Omega)=\prod_{(n,m)\in {\bf Z}^{2h}\backslash{(0,0)}}
\lambda_{n,m},
\label{llkd}\end{equation}
and
\begin{equation}
\widetilde{\cal Z}_h(\Omega)=\prod_{(n,m)\in {\bf Z}^{2h}\backslash{(0,0)}}
\mu_{n,m},
\label{0l1l2k5d4}\end{equation}
that on the torus reduce to the determinant of the Laplacian
\begin{equation}
{\cal Z}_1(\tau)=\widetilde{\cal Z}_1(\tau)= {\tau^{(2)}}^2|\eta(\tau)|^4.
\label{llkdtorus}\end{equation}
It is possible to get some insight on ${\cal Z}_h(\Omega)$ 
and $\widetilde{\cal Z}_h(\Omega)$ by 
investigating the behaviour of the $\lambda_{n,m}$'s and $\mu_{n,m}$'s
under pinching of the separating and non-separating
cycles of $\Sigma$. This can be done because the behaviour
of the period matrix near the boundary of the
moduli space is well-known. In particular the structure of
the $\lambda_{n,m}$'s and $\mu_{n,m}$'s seems to be suitable 
to recover the eigenvalues
of the Laplacian on the torus in the ``first'' component 
of the boundary of the compactified moduli space.
However we do not perform
such analysis here.

Another possible investigation concerning the results in this section
is the analysis of the subspace of the differentials 
in ${\cal T}^{p,q}$ made up of the scalar functions
$h_{n,m}$, $\overline h_{n,m}$ suitably 
combined with products of the $KN$ differentials 
$\psi_k^{(p)}$ and $\overline\psi_l^{(q)}$.

Going back to the construction of functions 
in ${\cal C}_+^\infty$ we notice that,
considering the functions in the 
set ${\cal F}$
as Fourier modes on higher genus
Riemann surfaces,
the Liouville field can be expanded as 
\begin{equation}
\sigma(z,\bar z)=\sum_{({n,m})\in {\bf Z}^{2h}} a_{n,m} 
{f_{n,m}\over \overline f_{n,m}}, \qquad \overline a_{n,m}=a_{-n,-m}.
\label{aaweylfactor}\end{equation}

\mysection{Liouville Action And Topological Gravity}

In this section we show that the classical Liouville action appears in the 
intersection numbers on moduli space. These numbers are the correlators of 
topological gravity as formulated by Witten \cite{1,2}. This result provides 
an explicit  relation between topological and Liouville gravity.

\subsection{Compactified Moduli Space}

We now introduce the moduli space of stable curves
$\overline{\cal M}_{h}$, that is the Deligne-Mumford 
compactification of moduli space.
 $\overline{\cal M}_{h}$  is a projective
variety and its boundary
${D}=\overline{\cal M}_{h}\backslash {\cal M}_h$,
called the compactification
divisor, decomposes into a union of divisors
${D}_0,\ldots,{D}_{[h/2]}$ which are
 subvarieties of complex codimension one.
 
A Riemann surface $\Sigma$ belongs to
${D}_{k>0}\cong \overline{\cal M}_{h-k,1}\times
\overline{\cal M}_{k,1}$ if it
has one node separating it into two components of genus $k$
and $h-k$. The locus in ${D}_0\cong \overline{\cal M}_{h-1,2}$
consists of surfaces that become, on removal of the node,
genus $h-1$ double punctured surfaces. 
Surfaces with multiple nodes lie in the intersections
of the $D_k$.

The compactified moduli space $\overline{\cal M}_{h,n}$ of
Riemann surfaces with $n$-punctures $z_1,\ldots,z_n$
 is defined in an analogous way to $\overline{\cal M}_h$.
The important point now is
that the punctures never collide with the node.
Actually the configurations with $(z_i-z_j)\to 0$ are
stabilized by considering them as the limit in which the $n$-punctured
surface degenerates into a $(n-1)$-punctured surface and the three
punctured sphere.

Let us go back to the space $\overline{\cal M}_{h}$.
The divisors ${D}_k$ define cycles 
and thus classes in $H_{6h-8}({\overline{\cal M}}_h,{\bf Q})$.
It turns out that the components of $D$ together with
the divisor associated to $[\omega_{WP}]/2\pi^2$ provide a basis for
$H_{6h-8}({\overline{\cal M}}_h,{\bf Q})$.
The main steps to prove this are the following.
First of all recall that the Weil-Petersson K$\ddot{\rm a}$hler
form $\omega_{WP}$ extends as a closed form to
${\overline{\cal M}}_h$ \cite{wolpert0}, in particular \cite{wolpertis}
\begin{equation}
{[\omega_{WP}]\over 2\pi^2}\in H^2({\overline{\cal M}}_h,{\bf Q}),
\label{ddcj}\end{equation}
which by Poincar\'e duality defines a cycle $D_{WP}/2\pi^2$ in
$H_{6h-8}({\overline{\cal M}}_h,{\bf Q})$.
The next step is due to Harer \cite{harer} who proved that
$H_2({\cal M}_h,{\bf Q})={\bf Q}$ so that
by Mayer-Vietoris
\begin{equation}
H_2({\overline{\cal M}}_h,{\bf Q})=
{\bf Q}^{[h/2]+2}.\label{harerrs}\end{equation}
In \cite{wolpertis} Wolpert constructed a basis of
2-cycles $C_k$, $k=0,\ldots,[h/2]+1$ for
$H_2({\overline{\cal M}}_h,{\bf Q})$ and computed the intersection 
matrix 
\begin{equation}
A_{jk}=C_j\cdot D_k,\qquad 
j,k=0,\ldots,[h/2]+1,\label{poljm}\end{equation}
 where $D_{[h/2]+1}\equiv D_{WP}/2\pi^2$.
The crucial result in \cite{wolpertis} is that $A_{jk}$ is not
singular so that the classes associated to $D_k$, $k=0,\ldots,[h/2]+1$
are a basis for $H_{6h-8}({\overline{\cal M}}_h,{\bf Q})$.

 Let us now define the universal curve
${\cal C}\overline{\cal M}_{h,n}$
 over $\overline{\cal M}_{h,n}$. It
  is built by placing over each point of
$\overline{\cal M}_{h,n}$ the Riemann surface
 which that point denotes.
Of course $\overline{\cal M}_{h,1}$ can be identified
with ${\cal C}\overline{\cal M}_{h}$. More generally
$\overline{\cal M}_{h,n}$ can be identified with 
${\cal C}_n\left(\overline {\cal M}_h\right)\backslash\{sing\}$ 
where ${\cal C}_n\left(\overline {\cal M}_h\right)$ denotes
 the $n$-fold fiber product of the
$n$-copies ${\cal C}_{(1)}\overline{\cal M}_h,
\ldots,{\cal C}_{(n)}\overline{\cal M}_h$
of the universal curve over $\overline{\cal M}_h$ and $\{sing\}$ is the 
locus of ${\cal C}_n\left(\overline {\cal M}_h\right)$ where the punctures 
come together.

Finally we define  $K_{{\cal C}/{\cal M}}$ as the cotangent
 bundle to the fibers of ${\cal C}\overline {\cal M}_{h,n}
\to\overline{\cal M}_{h,n}$,
it is built by taking all the spaces of
$(1,0)$-forms on the various $\Sigma$ and pasting
them together into a bundle over ${\cal C}
\overline{\cal M}_{h,n}$.

\subsection{$\big <\kappa_{d_1-1}\cdots\kappa_{d_n-1}\big >$}

 Let $\Sigma$ be a Riemann surface in $\overline{\cal M}_{h,n}$.
The cotangent space $T^*\Sigma_{|_{z_i}}$ varies
holomorphically with $z_i$ giving a holomorphic line bundle
${\cal L}_{(i)}$ on $\overline{\cal M}_{h,n}$.
Considering the $z_i$ as sections of the universal
 curve ${\cal C}\overline{\cal M}_{h,n}$
 we have  ${\cal L}_{(i)}=z_i^*\left(K_{{\cal C}/
{\cal M}}\right)$.

Let us consider the intersection numbers \cite{1,2}
 \begin{equation}
\big<\tau_{d_1}\cdots\tau_{d_n}\big>=
\int_{\overline{\cal M}_{h,n}}
c_1\left({\cal L}_{(1)}\right)^{d_1}\wedge\cdots
\wedge c_1\left({\cal L}_{(n)}
\right)^{d_n},\label{43}\end{equation}
where the power $d_i$ denotes the $d_i$-fold wedge
product. Notice that, since $c_1\left({\cal L}_{(i)}\right)$
is a two-form, $\big <\tau_{d_1}\cdots\tau_{d_n}\big >$
 does not depend on the ordering and, by
 dimensional arguments,
it may be nonvanishing only if the charge conservation
condition $\sum d_i=3h-3+n$ is satisfied. 
Moreover, due to the orbifold
nature of ${\cal M}_{h,n}$, the intersection numbers will generally 
be rational.

Related to the  $\tau$'s there are the so-called
Mumford tautological classes \cite{mumford1}.
Let $\pi : \overline {\cal M}_{h,1}\to\overline{\cal M}_h$
be the projection forgetting the puncture. The  tautological
classes are
\begin{equation}
\kappa_l=\pi_*\left(c_1\left({\cal L}\right)^{l+1}\right)
 =\int_{\pi^{-1}(p)}c_1\left({\cal L}\right)^{l+1},\qquad l\in {\bf Z}^+,
 \; p \in \overline {\cal M}_h,
\label{44}\end{equation}
where $\cal L$ is the line bundle whose fiber is the
cotangent space to the one marked point of
$\overline{\cal M}_{h,1}$.
 The $\kappa$'s correlation functions
are $\big <\kappa_{s_1}\cdots\kappa_{s_n}\big >=
\big <\wedge_{i=1}^n\kappa_{s_i},\overline{\cal M}_h \big >$.
To get the charge conservation condition we must take into account the
fact that integration on the fibre in (\ref{44}) decreases one (complex) 
dimension so that $\kappa_l$ is a $(l,l)$-form on the moduli space.
It follows that the nonvanishing condition for the intersection numbers
$\big <\kappa_{s_1}\cdots\kappa_{s_n}\big >$ is $\sum_i s_i=3h-3$.

There are relationships between the $\kappa$'s and $\tau$'s correlators.
 For example performing the integral over the fiber
of $\pi:\overline{\cal M}_{h,1}\to\overline{\cal M}_h$,
 we have 
\begin{equation}
\big <\tau_{3h-2} \big >= 
\int_{\overline{\cal M}_{h,1}}c_1({\cal L})^{3h-2}=
\int_{\overline{\cal M}_{h}}\kappa_{3h-3}=
\big <\kappa_{3h-3}\big >.
\label{dhadlk}\end{equation}
To find the general relationships between the $\kappa$'s and $\tau$'s 
correlators it is useful to write $\big <\kappa_{s_1}\cdots
\kappa_{s_n}\big >$ in the following way \cite{2}
 \begin{equation}
\big <\kappa_{d_1-1}\cdots\kappa_{d_n-1}\big >=
\int_{{\cal C}_n\left(\overline {\cal M}_h\right)}c_1
\left(\hat{\cal L}_{(1)}\right)^{d_1}
\wedge\cdots\wedge c_1\left(\hat{\cal L}_{(n)}
\right)^{d_n},\label{45}\end{equation}
where $\hat{\cal L}_{(i)}=\pi_i^*\left(K_{{\cal C}_{(i)}/
{\cal M}}\right)$  and  $\pi_i:{\cal C}_n\left(\overline {\cal M}_h\right)
\to{\cal C}_{(i)}\overline{\cal M}_h$ is the natural projection.
Then notice that ${\cal C}_n\left(\overline {\cal M}_h\right)$
and $\overline {\cal M}_{h,n}$ differ for a divisor at infinity only.
This is the unique difference between
$\big <\kappa_{d_1-1}\cdots\kappa_{d_n-1}
\big >$ and $\big<\tau_{d_1}\cdots\tau_{d_n}\big>$ as defined
in (\ref{43}) and (\ref{45}). Thus it is
possible to get  relations for arbitrary correlators.

\subsection{$\kappa_1={i\over 2\pi^2}\overline\partial\partial S_{cl}^{(h)}$}

We now show how
the scalar Laplacian defined with respect to 
the Poincar\'e metric enters in the expression of
the first tautological class on the moduli space. In order to do this
we  first introduce the determinant
line bundles on
the moduli space ${\cal M}_h$
\begin{equation}
\lambda_n=\det \,{\rm ind}\,\overline \partial_n.
\label{dtrmntlnb}\end{equation}
They are the maximum wedge powers of the space of holomorphic 
$n$-differentials. The line bundles $\lambda_1(\equiv \lambda_H)$ 
and $\lambda_2$ are the Hodge and the canonical line bundles respectively. 

In \cite{wolpert3} it has been shown that $\kappa_1=\omega_{WP}/\pi^2$ thus, 
by standard results on $\omega_{WP}$, we have
\begin{equation}
\kappa_1={6 i\over \pi }\overline\partial\partial \log {\det\, 
\Omega^{(2)}\over \det' \Delta_{\hat g,0}},\label{ftclss}\end{equation}
where $\partial$, 
 $\overline \partial$ denote the holomorphic and antiholomorphic 
components of the external derivative $d=\partial+\overline\partial$
on the moduli space.
Therefore the first tautological class can be seen as the curvature
form (that is $\kappa_1=12 c_1(\lambda_H)$ in ${\cal M}_h$)
of the Hodge line bundle $(\lambda_H;\langle \,,\rangle_Q)$
endowed with the Quillen norm
\begin{equation}
\langle \omega\,,\omega\rangle_Q=
{\det\,\Omega^{(2)}\over \det' \Delta_{\hat g,0}},\qquad
\omega=\omega_1\wedge \ldots\wedge \omega_h.\label{quillen}\end{equation}

As we have seen the  Liouville action  (\ref{la1}) evaluated
on the classical solution is a potential 
of $\omega_{WP}$ projected onto the Schottky space.
Thus in this space
\begin{equation}
\kappa_1={i\over 2\pi^2}\overline\partial\partial S_{cl}^{(h)},
\label{newtr1}\end{equation}
which provides a direct link between Liouville and topological gravity.

\mysection{Cutoff In 2D Gravity And The Background Metric}

Here we apply classical results on univalent (schlicht) functions in order to 
derive an inequality involving the cutoff of 2D gravity and the background 
geometry.

\subsection{Background Dependence In The Definition Of The Quantum Field}

An important aspect arising in quantum gravity is the
problem of the choice of the cutoff. In 2D quantum gravity
a related problem appears when we consider the norm of the 
Liouville field $\sigma$ defined by 
\begin{equation}
g=e^{\sigma} \hat g,\label{wl}\end{equation}
with $\hat g$ a background metric that we suppose to be in the conformal 
form $ds^2=2\hat g_{z\bar z}|dz|^2$.
The choice of the background  is an important step as
it defines the classical solution. 
To explain this point more thoroughly,
we first consider the relationship between the scalar curvatures
\begin{equation}
\sqrt {\hat g}\Delta_{\hat g,0}\sigma=\sqrt g R_g-\sqrt {\hat g}R_{\hat g},
\label{reltss}\end{equation}
where
\begin{equation}
\Delta_{\hat g,0}=-{\hat g}^{z\bar z}\partial_z\partial_{\bar z},
\label{lplcn}\end{equation}
is the scalar Laplacian for the conformal metric.
When $R_g=cst<0$, $\sigma=\sigma_{cl}$ of eq.(\ref{reltss}) 
is the solution of the classical equation of motion 
defined by the Liouville action
in the background metric $\hat g$. Thus both the solution of the  
equation of motion and the splitting
\begin{equation}
\sigma=\sigma_{cl}[\hat g]+\sigma_{qu}[\hat g],\label{spltng}\end{equation}
are background dependent. The
background dependence appears
in the path-integral formulation of Liouville gravity where the 
measure ${\cal D}_{\hat g}\sigma$, defined by the scalar product
\begin{equation}
||\delta \sigma ||^2_{\hat g}=\int_\Sigma \sqrt{\hat g} e^\sigma
 |\delta \sigma|^2,
\label{nontrasl}\end{equation}
is not translationally invariant.

Let us now choose  the Poincar\'e metric as background
\begin{equation}
\hat g_{z\bar z}={e^\varphi\over 2},\label{pncr3}\end{equation}
where $\varphi$ is given in (\ref{2}).
Before investigating its role in defining the quantum cutoff let us
notice that since
\begin{equation}
\sigma_{cl}[\hat g=e^{\varphi}]=0,\label{zrcl}\end{equation}
the $\sigma$ field in (\ref{spltng}) reduces to a full quantum field
and the Liouville action for $\sigma$, written with respect
to the background metric $\hat g$ and  evaluated
on the classical solution, reduces to the 
area of $\hat g$ which is just the topological number
$-2\pi \chi(\Sigma)$. 

\subsection{The Cutoff In $z$-Space}

In \cite{ddk} it was conjectured
that the Jacobian that arises in using the translation invariant measure
\begin{equation}
||\delta \sigma ||^2=\int_\Sigma \sqrt{\hat g}  |\delta \sigma|^2,
\label{trasl}\end{equation}
is given by the exponential of the Liouville action with modified 
coefficients.
Arguments in support of this conjecture may be found in \cite{mmdk}.
Some aspects of this conjecture are related to
the choice of the regulator.
We now show how the choice of the Poincar\'e metric
$d\hat s^2=e^\varphi |dz|^2$  as background
makes it possible to find an inequality involving the quantum cutoff and
classical geometry. Including the Liouville 
field $\sigma$ we have 
\begin{equation}
ds^2=e^{\phi}|dz|^2, \qquad \phi=\sigma+\varphi,\qquad 
\sigma=\sigma_{cl}[e^\varphi]+\sigma_{qu}[e^\varphi]=\sigma_{qu}[e^\varphi].
\label{wyl}\end{equation}
It is well-known that
 the cutoff in $z$ space $(\Delta z)^2_{min}$ is $z$-dependent 
\begin{equation}
(\Delta s)^2_{min}=
e^\phi(\Delta z)^2_{min}=\epsilon,\label{ctf}\end{equation} 
that is 
\begin{equation}
(\Delta z)^2\ge \epsilon e^{-\sigma-\varphi}.\label{dctff}\end{equation}

We stress that the cutoff arises already at the classical level.
As an example we consider a Riemann sphere with $n\ge 3$
punctures (we choose the standard normalization 
$z_{n-2}=0,z_{n-1}=1$ and 
$z_n=\infty$)
\begin{equation}
\Sigma= {\bf C}\backslash\{z_1,\ldots ,z_{n-3},0,1\}.
\label{rsphss}\end{equation}
  Near a puncture the
Poincar\'e metric has the following behaviour
\begin{equation}
\varphi(z) \sim
-2{\log}|z-z'|-2 {\log}|{\log}|z-z'||.
\label{rfdf}\end{equation}
Note that $e^\varphi$ is well-defined on the punctured surface:
to deleting the point provides a sort of
``topological'' cutoff for $\varphi$ which is related to the 
univalence of the inverse map of uniformization. 

The topological cutoff is related to the covariance of the Poincar\'e 
metric. To understand this it is instructive to write down the Liouville 
action on the Riemann spheres with $n$-punctures \cite{0}
\begin{equation}
S^{(0,n)}=\lim_{r\to 0}S^{(0,n)}_r=
\lim_{r\to 0}\left[\int_{\Sigma_r}
\left(\partial_z\phi\partial_{\bar z}{\phi}+e^{ \phi}\right)+
2\pi (n {\log} r+2(n-2){\log}|{\log}r|)\right],
\label{32}\end{equation}
$$
\Sigma_r=\Sigma\backslash\left(\bigcup_{i=1}^{n-1}
\{z||z-z_i|<r\}\cup\{z||z|>r^{-1}\}\right),
$$
where the field $\phi$ is in the class of smooth functions on
$\Sigma$ with the boundary condition
given by the asymptotic behaviour (\ref{rfdf}).
 Eq.(\ref{32}) shows that already at the classical level
the Liouville action needs a regularization 
whose effect is to cancel the contributions
coming from the non covariance of $|\phi_z|^2\notin{\cal T}^{1,1}$
and provides a modular anomaly for the 
Liouville action which  is strictly
related to the geometry of the moduli space \cite{asym}.
This classical geometric context 
is the natural framework 
to understand the relationships 
between covariance, regularization and modular anomaly.
In particular the relation between regularization and conformal weight
in this framework is similar to the analogous relation which 
arises in CFT where the scaling behaviour is fixed by normal ordering
and regularization. The fact that classical Liouville theory 
encodes a quantum feature such as regularization may be connected to the
fact that for the canonical 
transformation relating a particle moving in a Liouville potential
to a free particle, the effective quantum generating function is 
identical to its classical counterpart \cite{gha} (no normal ordering
problems).  
Furthermore, as we have seen, the 
link between classical Liouville theory and normal ordering appears also
in the analysis of cocycles.

\subsection{Univalent Functions And $(\Delta z)^2_{min}$}

The correlators of the one dimensional string 
have the structure (see \cite{poly} for notation)
\begin{equation}
G(p_1,\ldots,p_N)=F(p_1,\ldots,p_N)+ {A(p_1,\ldots,p_N)\over \sum 
\epsilon(p_i)+2b},\label{poliytu}\end{equation}
where the reason for the denominator, instead of the usual 
delta-function, is that the Liouville mode 
represents a positively defined metric.
 It seems that the boundaries
of the space of the ``half-infinite'' configuration space are related to 
the inequalities that naturally appear in the theory of univalent 
functions and in particular to their role in the 
uniformization theory which, as it has been shown in
 \cite{0}, is strictly related 
to classical Liouville theory.

Let us consider a simply connected domain $D$ of $\widehat{\bf C}$ with more 
than one boundary point. The Poincar\'e metric on $D$ reads
\begin{equation}
e^{\varphi_D(z,\bar z)}=
4{|f'_D(z)|^2\over (1-|f_D(z)|^2)^2},\label{poid}\end{equation}
where $f_D: D\to\Delta$ is a conformal mapping. 
We are interested in the bounds of $e^{\varphi_D}$. By an application
of the Schwarz lemma it can be proved that 
\begin{equation}
e^{\varphi_D(z,\bar z)}(\Delta_Dz)^2 \le 4,\label{nqltg}\end{equation} 
where $\Delta_D z$ denotes the Euclidean distance between $z$ and
the boundary $\partial D$. 
The lower bound is 
\begin{equation}
e^{\varphi_D(z,\bar z)}(\Delta_D z)^2 \ge 1,
\label{a01}\end{equation}
where now it is assumed that $\infty\notin D$. 
By eq.(\ref{usty}) we can express the metric in terms of the 
Schr\"odinger wave 
functions satisfying (\ref{new1}) so that (\ref{a01}) reads
\begin{equation}
\Delta_D z\ge \psi\overline \psi {\int \psi^{-2} -
\int \overline\psi^{-2}\over 2i}.\label{a02}\end{equation}
Eq.(\ref{a01}) follows from the 
Koebe {\it one-quarter theorem} \cite{lehto} stating that the boundary of the 
map of $|z|<1$ by any univalent and holomorphic function $f$ is always
at an Euclidean distance not less than $1/4$ from $f=0$.
Thus if $D$ is the unit disc and $f(0)=0$ we have $|f(z)|\ge 1/4$.
We stress that (\ref{a01}) resembles a sort of geometric uncertainty 
relation. 
In particular
 (\ref{nqltg}) and  (\ref{a01}) can be considered as infrared and ultraviolet
cutoff respectively.

Let us now consider the cutoff on $D$. By
(\ref{dctff}) and (\ref{nqltg}) it follows that 
\begin{equation}
(\Delta z)^2\ge {\epsilon\over 4} e^{-\sigma_D}(\Delta_D z)^2,
\label{nqltcq}\end{equation}
which relates the quantum  cutoff to the background geometry.

A related result concerns the Nehari theorem \cite{nehari}. 
It states that a sufficient condition for the univalence of 
a function $g$ is
\begin{equation}
e^{-\psi}|\{g,z\}|\le 2,\qquad |z|<1,
\label{neharith}\end{equation}
whereas the necessary condition is 
\begin{equation}
e^{-\psi}|\{g,z\}|\le 6,\qquad |z|<1,
\label{necessa}\end{equation}
where $e^{\psi}=(1-|z|^2)^{-2}$.
It can be shown that the constant 2 in (\ref{neharith}) cannot be
replaced by any larger one. Eqs.(\ref{neharith},\ref{necessa})
are inequalities between the 
Poincar\'e metric and the modulus of the Schwarzian derivative
of a univalent function which is related to the stress tensor.

\vspace{0.5cm}

{\bf Acknowledgements}:
I would like to thank G. Bonelli, J. Gibbons and G. Wilson
for stimulating discussions.
This work
has been partly supported by a SERC fellowship.

\end{document}